\documentclass{article}
\usepackage{amssymb}
\usepackage{amsmath}
\numberwithin{equation}{section} 
 \newcount\timehh\newcount\timemm
 \timehh=\time \divide\timehh by 60 \timemm=\time \count255=\timehh\multiply\count255 by -60
 \advance\timemm by \count255
 \DeclareMathOperator{\sgn}{\rm sgn}

 \DeclareMathOperator*{\res}{\rm res}
 
 \DeclareMathOperator{\R}{\mathbb{R}}
 \DeclareMathOperator{\C}{\mathbb{C}}
 \DeclareMathOperator{\Ss}{\mathcal{S}}

 \DeclareMathOperator{\Lo}{\mathcal{L}}
 \DeclareMathOperator{\Ro}{\mathcal{R}}
 
 \DeclareMathOperator{\Fo}{\mathcal{F}}
 \DeclareMathOperator{\q}{\mathbf{q}}
 \DeclareMathOperator{\bk}{\mathbf{k}}
 
 \DeclareMathOperator{\kap}{\varkappa}

\begin{document}
\title{Towards spectral theory of the Nonstationary Schr\"{o}dinger
equation with a two-dimensionally perturbed one-dimensional
potential\footnote{This work is supported in part by the Russian
Foundation for Basic Research (grant \# 02-01-00484) and by PRIN
2002 `Sintesi'.}}
\author{M.~Boiti$^{a}$, F.~Pempinelli$^{a}$, A.K.~Pogrebkov$^{b}$, B.~Prinari$^{a}$ \\
{\small $^{a}$ Dipartimento di Fisica, Universit\`a di Lecce and
Sezione INFN, Lecce, Italy} \\
{\small $^{b}$ Steklov Mathematical Institute, Moscow, Russia} \\
{\small email: boiti@le.infn.it, pempi@le.infn.it,
pogreb@mi.ras.ru, prinari@le.infn.it}}
\date{}
\maketitle
\begin{abstract}
The Nonstationary Schr\"{o}dinger equation with potential being a
perturbation of a generic one-dimensional potential by means of a
decaying two-dimensional function is considered here in the
framework of the extended resolvent approach. The properties of
the Jost solutions and spectral data are investigated.
\end{abstract}
\section{Introduction}

The Kadomtsev--Petviashvili equation in its version called KPI
~\cite{bib10}--\cite{bib12}
\begin{equation}
(u_{t}-6uu_{x_{1}}+u_{x_{1}x_{1}x_{1}})_{x_{1}}=3 u_{x_{2}x_{2}},
\label{KPI}
\end{equation}
is a (2+1)-dimensional generalization of the celebrated
Korteweg--de~Vries (KdV) equation. As a consequence, the KPI
equation admit solutions that behave at space infinity like the
corresponding solutions of the KdV equation. For instance, if
$u_{1}(t,x_{1})$ obeys KdV, then
$u(t,x_{1},x_{2})=u_{1}(t,x_{1}+\mu x_{2}+3\mu^{2}t)$ solves KPI
for an arbitrary constant $\mu \in \mathbb{R} $. Thus, it is
natural to consider solutions of~(\ref{KPI}) that are not decaying
in all directions at space infinity but have 1-dimensional rays
with behavior of the type of $u_{1}$. Without loss of generality,
one can restrict to taking $\mu =0$ (the generic case is
reconstructed by means of the Galileo invariance of~(\ref{KPI}).)

Even though KPI has been known to be integrable for about three
decades~\cite{bib11,bib12}, the general theory is far from being
complete. Indeed, the Cauchy problem for KPI with rapidly decaying
initial data was first studied by using the Inverse Scattering
Transform (IST) method in~\cite{bib13}-\cite{bib16}, the
associated spectral operator being the nonstationary
Schr\"{o}dinger operator
\begin{equation}
\mathcal{L}(x,\partial _{x}^{})=i\partial _{x_{2}}^{}+\partial
_{x_{1}}^{2}-u(x),\qquad x=(x_{1}^{},x_{2}^{}).  \label{NS}
\end{equation}
However, it is known that the standard approach to the spectral
theory of the operator~(\ref{NS}), based on integral equations for
the Jost solutions, fails for potentials with one-dimensional
asymptotic behavior.

In~\cite{bib1}-\cite{bib9} the method of the extend resolvent was
suggested as a way of pursuing a generalization of the IST that
enables considering operators with nontrivial asymptotic behavior
at space infinity.

The spectral theory for the simplest example of such potentials
in~(\ref{NS}) was developed in~\cite{bib17}, where the Cauchy
problem for the KPI equation was considered with initial data
\begin{equation}
u(x)=u_{1}^{}(x_{1}^{})+u_{2}^{}(x),  \label{potential}
\end{equation}
$u_{1}(x_{1})$ being the value at time $t=0$ of the one-soliton
solution of the KdV equation and $u_{2}(x)$ a smooth, real and
rapidly decaying function on the $(x_{1},x_{2})$-plane.
In~\cite{bib17} the direct problem was studied by using a modified
integral equation and it was shown that the modified Jost
solution, in addition to the standard jump across the real
$\mathbf{k}$-axis, where $\mathbf{k}$ is the spectral parameter,
also has a jump across a segment of the imaginary axis of the
complex $\mathbf{k}$-plane. However, some essential properties of
the Jost solutions and relations between spectral data were
missing. In~\cite{bib18}, in the framework of the extended
resolvent approach, the problem was completely solved for the case
where $u_{1}$ is a pure one-soliton potential. Here we extend the
results to the case of generic (smooth and rapidly decaying)
one-dimensional potential $u_{1}$. Some results are given here
without proofs, which will be provided in a forthcoming
publication.

\section{Basic objects of the resolvent approach}

Here we briefly review the basic elements of the extended
resolvent
approach. For further details, we refer the interested readers to \cite{bib1}%
-\cite{bib9}.

Let $\mathcal{A}=\mathcal{A}(x,\partial _{x})$ denote a
differential operator with kernel $A(x,x^{\prime
})=\mathcal{A}(x,\partial _{x}^{{}})\delta (x-x^{\prime })$,
$\delta (x)=\delta (x_{1})\delta (x_{2})$ being the
two-dimensional $\delta $-function. In what follows we consider
differential operators whose kernels $A(x,x^{\prime })$ belong to the space $%
\mathcal{S}^{\prime }$ of tempered distributions of the four real variables $%
x$ and $x^{\prime }$. We introduce an \textbf{extension} of
differential operators, i.e., to any differential operator
$\mathcal{A}$ we associate the differential operator $A(q)$ with
kernel
\begin{equation}
A(x,x^{\prime };q)\equiv e_{{}}^{-q(x-x^{\prime })}A(x,x^{\prime })=\mathcal{%
A}(x,\partial _{x}^{{}}+q)\delta (x-x^{\prime }),  \label{2}
\end{equation}
where $x$, $x^{\prime }$, $q\in \R^{2}$ and
$qx=q_{1}x_{1}+q_{2}x_{2}$. Such kernels form a subclass in the
space $ \mathcal{S}^{\prime }$ of tempered distributions of six
real variables. For generic elements $A(x,x^{\prime };q)$,
$B(x,x^{\prime };q)$ of this space $\mathcal{S}^{\prime }$, we
consider the standard composition rule
\begin{equation}
(AB)(x,x^{\prime };q)=\int dx^{\prime \prime }\,A(x,x^{\prime
\prime };q)\,B(x^{\prime \prime },x^{\prime };q).  \label{3}
\end{equation}
Since the kernels are distributions, this composition is neither
necessarily defined for all pairs of operators nor necessarily
associative.

An operator $A$ can have an inverse in the sense of the
composition law~(\ref{3}), say $AA^{-1}=I$ or $A^{-1}A=I$, where
$I$ is the unity operator in $\mathcal{S}^{\prime }$, $
I(x,x^{\prime };q)=\delta (x-x^{\prime }).$

The operation inverse to imbedding~(\ref{2}) consists in
associating to any operator $A(q)$, with kernel $A(x,x^{\prime
};q)$, its ``hat-kernel''
\begin{equation}
\widehat{A}(x,x^{\prime };q)=e_{{}}^{q(x-x^{\prime
})}A(x,x^{\prime };q). \label{7}
\end{equation}
In particular, for a differential operator $A$ the following
relations hold
\begin{equation}
(\widehat{AB})(x,x^{\prime };q)=\mathcal{A}(x,\partial
_{x}^{})\widehat{B} (x,x^{\prime };q),\quad
(\widehat{BA})(x,x^{\prime };q)=\mathcal{A}^{\text{d} }(x^{\prime
},\partial _{x^{\prime }}^{})\widehat{B}(x,x^{\prime };q),
\label{8.1}
\end{equation}
where $\mathcal{A}^{\text{d}}$ is the dual operator to
$\mathcal{A}$. In the following for equalities of the type
(\ref{8.1}) we shall use the notation
\begin{equation}
\widehat{AB}=\overrightarrow{\mathcal{A}}\widehat{B},\quad
\widehat{BA}= \widehat{B}\overleftarrow{\mathcal{A}}.  \label{8.2}
\end{equation}

\subsection{Resolvent approach in the case of rapidly decaying potential}

The operator extension $L(q)$ of $\mathcal{L}(x,\partial _{x})$
in~(\ref{NS} ) is given by
\begin{equation}
L(q)=L_{0}^{{}}(q)-U,  \label{9}
\end{equation}
where $L_{0}$ has kernel
\begin{equation}
L_{0}(x,x^{\prime };q)=\left[ i(\partial_{x_{2}}+q_{2})+(\partial
_{x_{1}}+q_{1})^{2}\right] \delta (x-x^{\prime }), \label{10}
\end{equation}
and $U$, which is called the potential operator, has kernel
\begin{equation}
U(x,x^{\prime };q)=u(x)\delta (x-x^{\prime }).  \label{12}
\end{equation}
Note that we will always assume $u(x)$ to be real. The main object
of our approach is the (extended) resolvent $M(q)$ of the operator
$L(q)$, which is defined as the inverse of the operator $L$, i.e.,
$M$ satisfies
\begin{equation}
LM=ML=I.  \label{14}
\end{equation}
The hat version (cf (\ref{7})) of the extended resolvent for the
``bare'' operator $L_{0}$ is given by
\begin{equation}
\widehat{M}_{0}(x,x^{\prime };q)=\frac{1}{2\pi
i}\int\limits_{\bk_{\Im }=q_{1}}d\bk_{\Re }\,\bigl[\theta
(x_{2}-x_{2}^{\prime })-\theta (\ell _{2\Im
}(\bk)-q_{2})\bigr]\,\Phi _{0}(x,\bk)\Psi _{0}(x^{\prime },\bk),
\label{16}
\end{equation}
(cf., for instance, \cite{bib18} for details) where we introduced
\begin{equation}
\Phi _{0}(x,\bk)=e^{-i\ell (\bk)x},\qquad \Psi _{0}(x,\bk%
)=e_{{}}^{i\ell (\bk)x},  \label{17}
\end{equation}
and the two component vector
\begin{equation}
\ell (\bk)=(\bk,\bk^{2}),\qquad \bk=\bk_{\Re }+i\bk_{\Im }\in \C.
\label{18}
\end{equation}
Note that we use boldface font to indicate that the corresponding
variables are complex valued. The functions $\Phi _{0}(x,\bk)$ and
$\Psi _{0}(x,\bk)$ solve the nonstationary Schr\"{o}dinger
equation~(\ref{NS}) and its dual in the case of zero potential,
and they can be considered as the Jost solutions for this trivial
case. Note also that they obey the conjugation property
\begin{equation}
\overline{\Phi _{0}^{{}}(x,\bk)}=\Psi _{0}^{{}}(x,\bar{\bk}).
\label{21}
\end{equation}
Using notation (\ref{8.2}), we can rewrite eqs. (\ref{14}) for the
case of zero potential in the form
$\overrightarrow{\mathcal{L}}_{0}^{{}}
\widehat{M}_{0}^{}(q)=\widehat{M}_{0}(q)\overleftarrow{\mathcal{L}}
_{0}^{{}}=I,$ which shows that $\widehat{M}_{0}(q)$ is a
two-parametric set of Green's functions of~(\ref{NS}) and its
dual.

From (\ref{16}) one directly obtains that for $q_{1}\ne0$
\begin{align}
\frac{\partial \widehat{M}_{0}^{{}}(q)}{\partial q_{1}^{{}}}&
=\frac{i}{\pi } \int\limits_{\bk_{\Im }=q_{1}}d\bk_{\Re
}^{{}}\,\bar{\bk}\delta (\ell _{2\Im }^{{}}(\bk)-q_{2}^{{}})\,\Phi
_{0}^{{}}(\bk)\otimes \Psi _{0}^{{}}(\bk),
\label{24} \\
\frac{\partial \widehat{M}_{0}^{{}}(q)}{\partial q_{2}^{{}}}&
=\frac{1}{2\pi i}\int\limits_{\bk_{\Im }=q_{1}}d\bk_{\Re
}^{{}}\,\delta (\ell _{2\Im }^{{}}( \bk)-q_{2}^{{}})\,\Phi
_{0}^{{}}(\bk)\otimes \Psi _{0}^{{}}(\bk),  \label{25}
\end{align}
where the direct product in~(\ref{24}), (\ref{25}) is defined in
the standard way as an operator with kernel
\begin{equation}
\left( \Phi _{0}(\bk)\otimes \Psi _{0}(\bk)\right) (x,x^{\prime
})=\Phi _{0}(x,\bk)\Psi _{0}(x^{\prime },\bk).  \label{26}
\end{equation}

\subsubsection{The resolvent and Hilbert identity}

The resolvent of the operator $L$ in (\ref{9}) with potential $u$
rapidly decaying can also be defined as the solution of the
integral equations
\begin{equation}
M=M_{0}^{{}}+M_{0}^{{}}UM,\qquad M=M_{0}^{{}}+MUM_{0}^{{}}.
\label{27}
\end{equation}
Under a small norm assumption on the potential we expect that the
solution $ M $ exists and is unique (the same for both integral
equations).

The resolvent satisfies the following analog of the Hilbert
identity
\begin{equation}
M^{\prime }-M=-M^{\prime }(L^{\prime }-L)M,  \label{29}
\end{equation}
where $L^{\prime }$ is another operator of the type~(\ref{NS}) and
$ M^{\prime }$ is its resolvent. Eq.~(\ref{29}) can be written in
the form
\begin{equation}
M(\q^{\prime })-M(\q)=M(\q^{\prime })L_{0}^{{}}(\q^{\prime
})(M_{0}^{{}}(\q ^{\prime })-M_{0}^{{}}(\q))L_{0}^{{}}(\q)M(\q),
\label{31}
\end{equation}
which can be used to obtain for the derivatives of the resolvent
$\widehat{M} (q)$ the following expressions
\begin{equation}
\frac{\partial \widehat{M}(q)}{\partial q_{j}^{{}}}=\widehat{M}(q)
\overleftarrow{\mathcal{L}}_{0}^{{}}\frac{\partial
\widehat{M}_{0}^{{}}(q)}{
\partial q_{j}^{{}}}\overrightarrow{\mathcal{L}}_{0}^{{}}\widehat{M}
(q),\quad j=1,2.  \label{32}
\end{equation}
Then by~(\ref{24}) and~(\ref{25}) for $q_{1}\neq 0$
\begin{align}
\frac{\partial \widehat{M}(q)}{\partial q_{1}^{{}}}=& \frac{i}{\pi
} \int\limits_{\bk_{\Im }=q_{1}}d\bk_{\Re }^{{}}\,\bar{\bk}\delta
(\ell _{2\Im
}^{{}}(\bk)-q_{2}^{{}})\,\Phi (\bk)\otimes \Psi (\bk),  \label{33} \\
\frac{\partial \widehat{M}(q)}{\partial q_{2}^{{}}}=&
\frac{1}{2\pi i} \int\limits_{\bk_{\Im }=q_{1}}d\bk_{\Re
}^{{}}\,\delta (\ell _{2\Im }^{{}}( \bk)-q_{2}^{{}})\,\Phi
(\bk)\otimes \Psi (\bk), \label{34}
\end{align}
where we introduced the functions
\begin{align}
& \Phi (x,\bk)=\int dx^{\prime
}\,\Bigl(\mathcal{L}_{0}^{\text{d}}(x^{\prime },\partial
_{x^{\prime }})\mathcal{G}(x,x^{\prime },\bk)\Bigr)\Phi
_{0}^{{}}(x^{\prime
},\bk),  \label{35} \\
& \Psi (x^{\prime },\bk)=\int dx\,\Psi
_{0}^{{}}(x,\bk)\,\mathcal{L} _{0}^{{}}(x,\partial
_{x})\,\mathcal{G}(x,x^{\prime },\bk), \label{36}
\end{align}
with $\mathcal{G}(x,x^{\prime },\bk)$ defined as a specific value
of the resolvent itself
\begin{equation}
\mathcal{G}(x,x^{\prime },\bk)=\widehat{M}(x,x^{\prime
};q)\Bigr|_{q=\ell _{\Im }(\bk )}^{{}}\equiv
\widehat{M}(x,x^{\prime };\bk_{\Im }^{{}},2\bk_{\Im }^{{}}\bk
_{\Re }^{{}}).  \label{37}
\end{equation}%
In what follows we consider the function $\mathcal{G}(x,x^{\prime
},\bk)$ as the kernel of the operator $\mathcal{G}(\bk)$ and the
functions $\Phi (x,\bk)$ and $\Psi (x^{\prime },\bk)$ as
``vector'' $\Phi (\bk)$ and ``covector'' $\Psi (\bk)$. For
shortness we write equations of the type~(\ref{35})--(\ref{37})
omitting the $x,x^{\prime }$-dependence, i.e. as
\begin{align}
& \Phi (\bk)=\mathcal{G}(\bk)\overleftarrow{\mathcal{L}}_{0}^{{}}\Phi _{0}^{{}}(\bk%
),\qquad \Psi (\bk)=\Psi
_{0}^{{}}(\bk)\overrightarrow{\mathcal{L}}
_{0}^{{}}\mathcal{G}(\bk),  \label{40} \\
& \mathcal{G}(\bk)=\widehat{M}(q)\Bigr|_{q=\ell _{\Im
}(\bk)}^{{}}.  \label{41}
\end{align}
In order to study the discontinuity at $q=0$ we introduce the
following notation for the specific limits of the resolvent at
this point of discontinuity
\begin{equation}
\mathcal{G}_{\pm }^{{}}(x,x^{\prime })=\lim_{q_{2}\rightarrow \pm
0}\lim_{q_{1}\rightarrow 0}\widehat{M}(x,x^{\prime };q),
\label{42}
\end{equation}
where the limit $q_{1}\rightarrow 0$ is independent of the sign.
Using again the Hilbert identity (\ref{31}), we get
\begin{equation}
\mathcal{G}_{+}^{{}}-\mathcal{G}_{-}^{{}}=\frac{1}{2\pi i}\int
dk\,\Phi _{\pm }^{{}}(k)\otimes \Psi _{\mp }^{{}}(k),  \label{46}
\end{equation}%
where we introduced the functions $\Phi _{\pm }^{{}}(x,k)$ and
$\Psi _{\pm }^{{}}(x,k)$ which, in analogy with~(\ref{40}), are
defined by
\begin{equation}
\Phi _{\pm }^{{}}(k)=\mathcal{G}_{\pm
}^{{}}\overleftarrow{\mathcal{L}}_{0}^{{}}\Phi _{0}^{{}}(k),\quad
\Psi _{\pm }^{{}}(k)=\Psi _{0}^{{}}(k)\overrightarrow{
\mathcal{L}}_{0}^{{}}\mathcal{G}_{\pm }^{{}}.  \label{47}
\end{equation}
Next we consider in details the properties of all the objects
introduced so far.

\subsubsection{Properties of the Green's function}

Thanks~(\ref{8.2}) it is clear that $\mathcal{G}(\bk)$ defined
in~(\ref{37}) is a Green's function of the operator $\Lo$
depending on the complex parameter $ \bk$, i.e.
$\overrightarrow{\Lo}\mathcal{G}(\bk)=\mathcal{G}(\bk)\overleftarrow{\Lo}=I.$
Also, due to the reality of $u(x)$, we have
\begin{equation}
\overline{\mathcal{G}(x,x^{\prime },\bk)}=\mathcal{G}(x^{\prime
},x,\overline{\bk}).  \label{49}
\end{equation}%
Applying the reduction~(\ref{37}) to equalities~(\ref{27}) it
follows that this function obeys the integral equations
\begin{equation}
\mathcal{G}(\bk)=\mathcal{G}_{0}^{{}}(\bk)+\mathcal{G}_{0}^{{}}(\bk)U\mathcal{G}(\bk),\qquad
\mathcal{G}(\bk)=\mathcal{G}_{0}^{{}}(\bk
)+\mathcal{G}(\bk)U\mathcal{G}_{0}^{{}}(\bk),  \label{50}
\end{equation}%
where the Green's function $\mathcal{G}_{0}(\mathbf{k})$ of the
operator $\mathcal{L} _{0}$ is defined by the general
formula~(\ref{41}) in terms of $M_{0}$. By~( \ref{50}) one can
check that $g(x,x^{\prime },\bk)=e_{{}}^{i\ell (\bk )(x-x^{\prime
})}\mathcal{G}(x,x^{\prime },\bk)$ is a bounded function of its
arguments and that $\lim_{\bk\rightarrow \infty }g(x,x^{\prime
},\bk)=0$ if the potential $u(x)$ decays rapidly enough.

The function $\mathcal{G}(x,x^{\prime },\bk)$ is a continuously
differentiable function of $\bk$ in the whole complex plane $\C$
with exception of the real axis $\bk_{\Im }=0$. By~(\ref{33})
and~(\ref{34}) for $\bk_{\Im }\neq 0$ we have
\begin{equation}
\dfrac{\partial \mathcal{G}(\bk)}{\partial \bk_{\Re
}^{{}}}=\dfrac{\sgn\bk_{\Im }^{{}} }{2\pi i}\,\Phi (\bk)\otimes
\Psi (\bk),\qquad \dfrac{\partial \mathcal{G}(\bk)}{
\partial \bk_{\Im }^{{}}}=\dfrac{\sgn\bk_{\Im }^{{}}}{2\pi }\,\Phi (\bk
)\otimes \Psi (\bk),  \label{54}
\end{equation}
so that in the complex domain this Green's function is analytic
for $\bk_{\Im }\neq 0$ and discontinuous at the real axis.

Properties of the functions $\mathcal{G}_{\pm }(x,x^{\prime })$ in
(\ref{42}) as well follow from the properties of the resolvent.
Both of them are also Green's functions of the operator~(\ref{NS})
and its dual, obey the conjugation property
\begin{equation}
\overline{\mathcal{G}_{\pm }^{{}}(x,x^{\prime })}=\mathcal{G}_{\mp
}^{{}}(x^{\prime },x), \label{57}
\end{equation}
and satisfy integral equations
\begin{equation}
\mathcal{G}_{\pm }^{{}}=\mathcal{G}_{0,\pm
}^{{}}+\mathcal{G}_{0,\pm }^{{}}U\mathcal{G}_{\pm }^{{}},\qquad
\mathcal{G}_{\pm }^{{}}=\mathcal{G}_{0,\pm }^{{}}+\mathcal{G}_{\pm
}^{{}}U\mathcal{G}_{0,\pm }^{{}}, \label{58}
\end{equation}
where $\mathcal{G}_{0,\pm }$ is given by%
\begin{equation}
\mathcal{G}_{0,\pm }^{{}}(x,x^{\prime })=\frac{\pm \theta (\pm
(x_{2}^{{}}-x_{2}^{\prime }))}{2\pi i}\int dk\,\Phi
_{0}^{{}}(x,k)\Psi _{0}^{{}}(x^{\prime },k).  \label{44}
\end{equation}
Since the resolvent is discontinuous at $q=0$, the limiting values
of the Green's function $\mathcal{G}(\bk)$ on the real axis,
$\mathcal{G}_{{}}^{\pm }(k)=\mathcal{G}(k\pm i0)$, do not
coincide, in general, with the Green's functions $\mathcal{G}_{\pm
}$. In order to find relations between them we again start
from~(\ref{31}) choosing both $ q^{\prime }$ and $q$ to be real.
Performing for $q^{\prime }$ and $q$ proper limiting procedures
one obtains
\begin{align}
\mathcal{G}_{{}}^{\sigma }(k)-\mathcal{G}_{\pm }^{{}}& =\frac{\mp
1}{2\pi i}\int dp\,\theta (\mp \sigma (k-p))\Phi _{\pm
}^{{}}(p)\otimes \left( \Psi _{0}^{{}}(p)
\overrightarrow{\mathcal{L}}_{0}^{{}}\mathcal{G}_{{}}^{\sigma
}(k)\right) ,  \label{64}
\\
\mathcal{G}_{{}}^{\sigma }(k)-\mathcal{G}_{\pm }^{{}}& =\frac{\mp
1}{2\pi i}\int dp\,\theta (\mp \sigma (k-p))\left(
\mathcal{G}_{{}}^{\sigma }(k)\overleftarrow{\mathcal{L}}
_{0}^{{}}\Phi _{0}^{{}}(p)\right) \otimes \Psi _{\pm
}^{{}}(p),\quad \sigma =+,-,  \label{65}
\end{align}%
where in the rhs we used a shorthand notation analogous to that
in~(\ref{40}).

\subsubsection{Jost and advanced/retarded solutions and bilinear
representation for the resolvent}

It is straightforward to check that the functions $\Phi (x,\bk)$
and $\Psi (x,\bk)$ defined by (\ref{40})\ obey the nonstationary
Schr\"{o}dinger equation with potential $u(x)$ and its dual, i.e.,
$\overrightarrow{\Lo}\Phi (\bk)=\Psi (\bk)\overleftarrow{\Lo}=0,$
and, thanks to (\ref{49}), they satisfy the conjugation property
\begin{equation}
\overline{\Phi (x,\bk)}=\Psi (x,\overline{\bk}).  \label{67}
\end{equation}%
The integral equations for these functions
\begin{equation}
\Phi (\bk)=\Phi _{0}^{{}}(\bk)+\mathcal{G}_{0}^{{}}(\bk)U\Phi
(\bk),\qquad \Psi (\bk )=\Psi _{0}^{{}}(\bk)+\Psi
(\bk)U\mathcal{G}_{0}^{{}}(\bk), \label{69}
\end{equation}%
follow from (\ref{50}). Finally, they obey the orthogonality
relation
\begin{equation}
\frac{1}{2\pi }\int dx_{1}^{{}}\,\Psi (x,\bk+p)\Phi (x,\bk)=\delta
(p),\qquad \bk\in \C,\quad p\in \mathbb{R}.  \label{75}
\end{equation}

The properties of the functions $\Phi _{\pm }(x,k)$ and $\Psi
_{\pm }(x,k)$ are derived analogously from their
definition~(\ref{47}). They also satisfy
$\overrightarrow{\Lo}\Phi_{\pm} (k)=\Psi_{\pm}
(k)\overleftarrow{\Lo}=0$, the integral equations
\begin{equation}
\Phi _{\pm }^{{}}(k)=\Phi _{0}^{{}}(k)+\mathcal{G}_{0,\pm
}^{{}}U\Phi _{\pm }^{{}}(k),\qquad \Psi _{\pm }^{{}}(k)=\Psi
_{0}^{{}}(k)+\Psi _{\pm }^{{}}(k)U\mathcal{G}_{0,\pm }^{{}},
\label{79}
\end{equation}
and the conjugation property
\begin{equation}
\overline{\Phi _{\pm }^{{}}(x,k)}=\Psi _{\mp }^{{}}(x,k).
\label{77}
\end{equation}
Also $\Phi _{\pm }(x,k)$ and $\Psi _{\pm }(x,k)$ satisfy an
orthogonality relation, i.e.,
\begin{equation}
\frac{1}{2\pi }\int dx_{1}^{{}}\,\Psi _{\pm }^{{}}(x,k+p)\Phi
_{\mp }^{{}}(x,k)=\delta (p),\qquad k,p\in \R.  \label{80}
\end{equation}

The properties of the Jost solutions enable us to reconstruct
$M(q)$ from~( \ref{33}) in the form
\begin{equation}
\widehat{M}(x,x^{\prime };q)=\frac{1}{2\pi i}\int\limits_{\bk_{\Im
}=q_{1}}d \bk_{\Re }^{{}}\,\bigl[\theta (x_{2}^{{}}-x_{2}^{\prime
})-\theta (\ell _{2\Im }^{{}}(\bk)-q_{2}^{{}})\bigr]\,\Phi
(x,\bk)\Psi (x^{\prime },\bk), \label{81}
\end{equation}
that generalizes~(\ref{16}) to the case of nonzero potentials.
Continuity of the resolvent at $q_{1}=0$ when $q_{2}\neq 0$
implies that
\begin{equation}
\int dk\,\Phi _{{}}^{+}(k)\otimes \Psi _{{}}^{+}(k)=\int dk\,\Phi
_{{}}^{-}(k)\otimes \Psi _{{}}^{-}(k)  \label{83}
\end{equation}
where we introduced for the limiting values of the Jost solutions
at the real axis the notation
\begin{equation}
\Phi _{{}}^{\pm }(x,k)=\Phi (x,k\pm i0),\qquad \Psi _{{}}^{\pm
}(x,k)=\Psi (x,k\pm i0).  \label{82}
\end{equation}

Representation~(\ref{81}) plays a crucial role in the resolvent
approach since it enables us to express all objects of the
spectral theory in terms of the Jost solutions. In particular,
thanks to~(\ref{37}), for the Green's function of the Jost
solutions we get
\begin{equation}
\mathcal{G}(x,x^{\prime },\bk)=\frac{1}{2\pi i}\int
dp\,\bigl[\theta (x_{2}^{{}}-x_{2}^{\prime })-\theta (\bk_{\Im
}^{{}}p)\bigr]\,\Phi (x,p+ \mathbf{k})\Psi (x^{\prime
},p+\mathbf{k}).  \label{84}
\end{equation}
Moreover, from~(\ref{81}) one can get the following expression of
the advanced/re\-tar\-ded Green's functions in terms of the
limiting values of the Jost solutions
\begin{equation}
\mathcal{G}_{\pm }^{{}}(x,x^{\prime })=\frac{\pm \theta (\pm
(x_{2}^{{}}-x_{2}^{\prime }))}{2\pi i}\int dk\,\Phi _{{}}^{\sigma
}(x,k)\Psi _{{}}^{\sigma }(x^{\prime },k),  \label{85}
\end{equation}
where $\sigma =+,-$ and we used notation~(\ref{82}). Note that
condition~( \ref{83}) guarantees that the $\mathcal{G}_{\pm }$'s
are independent of the sign $ \sigma $ in the rhs.

From the bilinear representation (\ref{81}) we also obtain the
following of completeness relation the Jost solutions
\begin{equation}
\frac{1}{2\pi }\int\limits_{x_{2}^{\prime }=x_{2}}d\bk_{\Re
}^{{}}\,\Psi (x^{\prime },\bk)\Phi (x,\bk)=\delta
(x_{1}-x_{1}^{\prime }).  \label{86}
\end{equation}

\subsubsection{Relations among Jost and advanced/retarded solutions.
Spectral data}

We obtained the bilinear representations (\ref{84}) and (\ref{85})
of the Green's functions for the Jost solutions of the Green's
function for the
advanced/retarded solutions. We derived also equations~(\ref{64}) and~(\ref%
{65}) relating these Green's functions. These relations can be
exploited for deriving relations among the advanced/retarded and
Jost solutions on the real axis and to use them to introduce the
scattering data. In fact, applying $\overleftarrow{\Lo}_{0}\Phi
_{0}(k)$ to~(\ref{64}) from the right and $\Psi
_{0}(k)\overrightarrow{\Lo}_{0}$ to~(\ref{65}) from the left,
recalling definitions~(\ref{40}) and~(\ref{47}), we get (notice
that $k\in \mathbb{R},\sigma =+,-,$)
\begin{equation}
\Phi _{{}}^{\sigma }(k)=\int dp\,\Phi _{\pm }^{{}}(p)r_{\pm
}^{\sigma }(p,k),\quad \Psi _{{}}^{\sigma }(k)=\int
dp\,\overline{r_{\mp }^{-\sigma }(p,k)}\Psi _{\pm }^{{}}(p),\qquad
\label{88}
\end{equation}
where
\begin{align}
& r_{\pm }^{\sigma }(p,k)=\delta (p-k)\mp \theta (\pm \sigma
(p-k))r_{{}}^{\sigma }(p,k),\quad p,k\in \R,  \label{89} \\
& r^{\sigma }(p,k)=\frac{\Psi
_{0}^{{}}(p)\overrightarrow{\mathcal{L}}
_{0}^{{}}\mathcal{G}_{{}}^{\sigma
}(k)\overleftarrow{\mathcal{L}}_{0}^{{}}\Phi _{0}^{{}}(k)}{2\pi
i}\equiv \frac{\Psi _{0}^{{}}(p)\overrightarrow{\mathcal{L
}}_{0}^{{}}\Phi _{{}}^{\sigma }(k)}{2\pi i},  \label{90}
\end{align}
are the scattering data. Recalling notations ~(\ref{35}),
(\ref{36}) and~(\ref{40}), the ``expectation values'' at the
numerator have the following explicit expressions
\begin{align}
& \Psi
_{0}^{{}}(p)\overrightarrow{\mathcal{L}}_{0}^{{}}\mathcal{G}_{{}}^{\sigma
}(k)
\overleftarrow{\mathcal{L}}_{0}^{{}}\Phi _{0}^{{}}(k)=  \nonumber \\
& \qquad =\int dx\int dx^{\prime }\,\Psi
_{0}^{{}}(x,p)\Bigl(\mathcal{L} _{0}^{{}}(x,\partial
_{x})\mathcal{L}_{0}^{\text{d}}(x^{\prime },\partial _{x^{\prime
}})\mathcal{G}_{{}}^{\sigma }(x,x^{\prime },k)\Bigr)\Phi
_{0}^{{}}(x^{\prime },k),  \label{91} \\
& \Psi _{0}^{{}}(p)\overrightarrow{\mathcal{L}}_{0}^{{}}\Phi
_{{}}^{\sigma }(k)=\int dx\,\Psi
_{0}^{{}}(x,p)\mathcal{L}_{0}^{{}}(x,\partial _{x})\Phi
_{{}}^{\sigma }(x,k),  \label{92}
\end{align}
showing that they are functions of $p$ and $k$ only.

In order to get the advanced/retarded solutions in terms of the
boundary values of the Jost ones we use~(\ref{84}) and the
limiting values of~(\ref{84}) on the real axis to obtain
\begin{equation}
\mathcal{G}_{{}}^{\sigma }(k)-\mathcal{G}_{\pm }^{{}}=\frac{\mp
1}{2\pi i}\int dp\,\theta (\pm \sigma (p-k))\Phi _{{}}^{\sigma
}(p)\otimes \Psi _{{}}^{\sigma }(p),\quad \sigma =+,-,  \label{94}
\end{equation}%
and in the same way as above we readily derive
\begin{equation}
\Phi _{\pm }^{{}}(k)=\int dp\,\Phi _{{}}^{\sigma
}(p)\overline{r_{\pm }^{-\sigma }(k,p)},\qquad \Psi _{\pm
}^{{}}(k)=\int dp\,r_{\mp }^{\sigma }(k,p)\Psi _{{}}^{\sigma }(p).
\label{96}
\end{equation}
Now inserting these expressions into (\ref{88}) and taking into
account the orthogonality properties~(\ref{75}) and~\eqref{80} of
the Jost and advanced/retarded solutions we derive that the
spectral data obey the following characterization
equations~\cite{bib19}
\begin{align}
& \int dp\,\overline{r_{\pm }^{-\sigma }(p,k)}r_{\pm }^{\sigma
}(p,k^{\prime
})=\delta (k-k^{\prime }),  \label{97} \\
& \int dp\,\overline{r_{\pm }^{-\sigma }(k^{\prime },p)}r_{\pm
}^{\sigma
}(k,p)=\delta (k-k^{\prime }),\quad \sigma =+,-,  \label{98} \\
& \int dp\,\overline{r_{+}^{\sigma }(p,k)}r_{+}^{\sigma
}(p,k^{\prime })=\int dp\,\overline{r_{-}^{\sigma
}(p,k)}r_{-}^{\sigma }(p,k^{\prime }). \label{99}
\end{align}

If we introduce the alternative scattering data
\begin{equation}
F_{{}}^{\sigma }(k,k^{\prime })=\int dp\,\overline{r_{\pm
}^{-\sigma }(p,k)} r_{\pm }^{-\sigma }(p,k^{\prime }), \label{101}
\end{equation}
we can express the discontinuity of the Jost solutions across the
real axis as follows
\begin{equation}
\Phi _{{}}^{\sigma }(k)=\int dp\,\Phi _{{}}^{-\sigma
}(p)F_{{}}^{-\sigma }(p,k),\qquad \Psi _{{}}^{\sigma }(k)=\int
dp\,F_{{}}^{\sigma }(k,p)\Psi _{{}}^{-\sigma }(p).  \label{102}
\end{equation}
Thanks to~(\ref{99}) these scattering data are independent of the
choice of the $+$ and $-$ sign in the r.h.s.\ of~(\ref{101}).
These scattering data satisfy the characterization equations
\begin{equation}
(F_{{}}^{\sigma })_{{}}^{\dagger }=F_{{}}^{\sigma },\qquad
F_{{}}^{-\sigma }=(F_{{}}^{\sigma })_{{}}^{-1}  \label{103}
\end{equation}
to which we have to add the requirement that $F_{{}}^{\sigma }$
can be
decomposed in the products~(\ref{101}) of two sets of triangular operators~(%
\ref{89}). Here again the first equation in~(\ref{103}) is related
to the reality requirement for $u(x)$ while the second equation is
a regularity requirement.

The inverse problem can be formulated in the standard way by using
the analyticity properties of the Jost solutions. In \cite{bib5}
we also demonstrated that the inverse problem can be formulated in
terms of the resolvent itself. Here we skip this for shortness, as
well as many other results that follows from the extended
resolvent approach.

\section{Case of one-dimensional potential}

\subsection{Resolvent approach}

Now we consider the imbedding of the standard one-dimensional
scattering transform in terms of two-dimensional differential
operator with one-dimensional potential. Specifically, we consider
the extended differential operator
\begin{equation}
L_{1}^{{}}(q)=L_{0}^{{}}(q)-U_{1}^{{}},\qquad
U_{1}^{{}}(x,x^{\prime };q)=u_{1}^{{}}(x_{1}^{{}})\delta
(x-x^{\prime }),  \label{135}
\end{equation}%
where $L_{0}^{{}}(q)$ is defined in~(\ref{10}). For the Jost
solution of the nonstationary Schr\"{o}dinger equation and its
dual we use notation
\begin{eqnarray}
&& \varphi (x,\bk)=e_{{}}^{-i\bk^{2}x_{2}}\Phi
_{1}^{{}}(x_{1},\bk)=e_{}^{-i\bk x_{1}-i\bk^{2}x_{2}}\chi_{1}^{}(x_{1},\bk%
),\\
&& \psi (x,\bk)=e_{{}}^{i\bk^{2}x_{2}}\Psi
_{1}^{{}}(x_{1},\bk)=e_{}^{i\bk x_{1}+i\bk^{2}x_{2}}\xi_{1}^{}(x_{1},\bk%
), \label{137}
\end{eqnarray}
where $\Phi _{1}(x_{1},k)$ and $\Psi _{1}(x_{1},k)$ are the Jost
solutions of the one-dimensional Sturm-Liouville operator, namely
$\chi_{1}$ is defined via the integral equation
\begin{equation}
\chi_{1}(x_{1},\bk)=1+\int\limits_{-\bk_{\Im}\infty}^{x_{1}}dy_{1}%
\dfrac{e^{2i\bk(x_{1}-y_{1})}-1}{2i\bk}u_{1}(y_{1})\chi_{1}(y_{1},\bk)
\label{107}
\end{equation}
and a similar equation defines the dual Jost solution. Functions
$\varphi (x,\bk)$ and $\psi (x,\bk)$, as well as their boundary
values at the real axis, obey a conjugation property
like~(\ref{67}), i.e.,
\begin{equation}
\overline{\varphi (x,\bk)}=\psi (x,\overline{\bk}),\qquad
\overline{\varphi ^{\pm }(x,k)}=\psi ^{\mp }(x,k),\quad k\in \R
\label{138:1}
\end{equation}
and satisfy orthogonality and completeness relations of the form
\begin{align}
& \frac{1}{2\pi }\int dx_{1}^{{}}\,\psi (x,\bk+p)\varphi
(x,\bk)=\dfrac{
\delta (p)}{t_{1}(\bk)},\quad p\in \R,\quad \bk\in \C,  \label{139} \\
& \int dx_{1}^{{}}\,\psi (x,i\varkappa _{j})\varphi (x,\bk)=\int
dx_{1}^{{}}\,\psi (x,\bk)\varphi (x,i\varkappa _{j})=0,\quad
|\bk_{\Im
}^{{}}|<\varkappa _{j},  \label{140} \\
& \int dx_{1}^{{}}\psi (x,i\varkappa _{j})\varphi (x,i\varkappa
_{j^{\prime
}})=\dfrac{i\delta _{j,j^{\prime }}}{t_{j}},  \label{141} \\
& \frac{1}{2\pi }\int\limits_{x_{2}^{\prime }=x_{2}^{{}}}d\bk_{\Re
}^{{}}\,t_{1}(\bk)\varphi (x,\bk)\psi (x^{\prime },\bk)-  \nonumber \\
& \qquad -i\sum_{j=1}^{N}t_{j}\theta (\varkappa _{j}-|\bk_{\Im
}^{{}}|)\varphi (x,i\varkappa _{j})\psi (x^{\prime },i\varkappa
_{j})\Bigr |_{x_{2}^{\prime }=x_{2}^{{}}}^{{}}=\delta
(x_{1}^{{}}-x_{1}^{\prime }) \label{142}
\end{align}%
where $t_{1}(\bk)$ is the transmission coefficient of the
one-dimensional problem, with poles at points $\pm i\varkappa
_{j}$ (for definiteness, we assume $\varkappa _{j}>0$ for all
$j=1,\dots N$), and residues at these points given by
\begin{equation}
t_{\pm j}=\res_{\bk=\pm i\varkappa _{j}}t_{1}(\bk).
\end{equation}
Another set of discrete spectral data for the one-dimensional
problem is given by the coefficients $b_{j},$ defined by one of
the following equalities
\begin{equation}
\Phi _{1}(x_{1},i\varkappa _{j})=b_{j}\Phi _{1}(x_{1},-i\varkappa
_{j}),\qquad \Psi _{1}(x_{1},-i\varkappa _{j})=b_{j}\Psi
_{1}(x_{1},i\varkappa _{j}).  \label{116}
\end{equation}
These coefficients are real and nonzero and such that
\begin{equation}
\sgn(it_{j}b_{j})=-1.  \label{124}
\end{equation}
By~\eqref{116} we have
\begin{equation}
\varphi (x,i\varkappa _{j})=b_{j}\varphi (x,-i\varkappa
_{j}),\qquad \varphi (x,i\varkappa _{j})=b_{j}\overline{\psi
(x,i\varkappa _{j})}.  \label{154}
\end{equation}

\subsection{Resolvent}

The resolvent of the one-dimensional $L$-operator that depends on
$\q_{2}$ as a parameter is given by
\begin{align}
\widehat{M}_{1}(x,x^{\prime };q)=& \dfrac{1}{2\pi
i}\int\limits_{\bk_{\Im }=q_{1}}d\bk_{\Re }\,[\theta
(x_{2}-x_{2}^{\prime })-\theta (2\bk_{\Re }\bk _{\Im
}-q_{2})]\,t_{1}(\bk)\varphi (x,\bk)\psi (x^{\prime },\bk)-
\nonumber
\\
& -\sum_{j}\theta (\varkappa _{j}^{2}-q_{1}^{2})t_{j}[\theta
(x_{2}-x_{2}^{\prime })-\theta (-q_{2})]\,\varphi (x,i\varkappa
_{j})\psi (x^{\prime },i\varkappa _{j}).  \label{162}
\end{align}

We decompose this kernel into the sum of a regular and a singular
parts, so that
\begin{equation}
\widehat{M}_{1}^{{}}(q)=\widehat{M}_{1,\text{reg}}^{{}}(q)+\sum_{j}\Gamma
_{j}(q)\,\varphi (i\varkappa _{j})\otimes \psi (i\varkappa _{j}),
\label{164}
\end{equation}
with $x$-independent functions
\begin{equation}
\Gamma _{j}(q)=\frac{t_{j}\sgn q_{1}}{2\pi i}\log \frac{
q_{2}^{{}}+2iq_{1}^{{}}(q_{1}^{{}}-\varkappa _{j})}{%
q_{2}^{{}}+2iq_{1}^{{}}(q_{1}^{{}}+\varkappa _{j})},\quad
j=1,\ldots ,N, \label{165}
\end{equation}
where we use for the logarithm the following definition
\begin{equation}
\log z=\log |z|+i\arctan \dfrac{z_{\Im }}{z_{\Re }}+i\pi \theta
(-z_{\Re }) \sgn z_{\Im }.  \label{166}
\end{equation}
With such definition, it follows that the extended resolvent in
the case of one-dimensional potential has logarithmic
singularities at the points $q=(\pm \varkappa _{j},0)$ with a cut
along $q_{2}=0$, since $\Gamma _{j}(q_{1},+0)-\Gamma
_{j}(q_{1},-0)=-t_{j}\theta (\varkappa _{j}-|q_{1}|)$.

\subsection{Properties of the resolvent and of the Green's function.}

For the discontinuity of the resolvent at $q_{2}=0$ we get
from~(\ref{162})
\begin{equation}
\widehat{M}_{1}(q_{1},+0)-\widehat{M}_{1}(q_{1},-0)=-\sum_{j}t_{j}\theta
(\varkappa _{j}-|q_{1}|)\,\varphi (i\varkappa _{j})\otimes \psi
(i\varkappa _{j}).  \label{168}
\end{equation}
For all other values of $q$ the resolvent~(\ref{162}) has
derivatives with respect to $q$ of the form of~(\ref{24}),
(\ref{25}). For instance
\begin{equation}
\frac{\partial \widehat{M}_{1}^{{}}(q)}{\partial
q_{2}^{{}}}=\frac{1}{2\pi i} \int\limits_{\bk_{\Im
}=q_{1}}d\bk_{\Re }^{{}}\,\delta (\ell _{2\Im }^{{}}(
\bk)-q_{2}^{{}})\,t_{1}(\bk)\varphi (\bk)\otimes \psi (\bk),
\label{169}
\end{equation}%
which at the vicinity of the point $q_{1}=0$ has to be considered
in the distributional sense. Thus, we can introduce the Green's
function $ \mathcal{G}_{1}(x,x^{\prime },\bk)$ of the Jost
solutions by using a reduction analogous to~(\ref{41}). Then,
from~(\ref{162}) we get the bilinear representation
\begin{align}
\mathcal{G}_{1}^{{}}(x,x^{\prime },\bk)=& \frac{1}{2\pi i}\int
d\alpha \,\bigl[\theta (x_{2}^{{}}-x_{2}^{\prime })-\theta
(\bk_{\Im }^{{}}(\alpha -\bk_{\Re }))
\bigr]\,t_{1}(\alpha +i\bk_{\Im })\times  \nonumber \\
& \times \varphi (x,\alpha +i\bk_{\Im })\psi (x^{\prime },\alpha
+i\bk_{\Im
})- \\
& -\sum_{j}t_{j}\theta (\varkappa _{j}-|\bk_{\Im
}^{{}}|)\,\bigl[\theta (x_{2}^{{}}-x_{2}^{\prime })-\theta
(-\bk_{\Re }^{{}}\bk_{\Im }^{{}})\bigr] \varphi (x,i\varkappa
_{j})\psi (x^{\prime },i\varkappa _{j}),  \label{170}
\end{align}
that generalizes to the case of generic potential $u_{1}$ the
Green's function used in~\cite{bib17} and~\cite{bib18} for pure
one-soliton potential. This Green's function obeys the conjugation
property~(\ref{49}) and the function $g_{1}(x,x^{\prime
},\bk)=e_{{}}^{i\ell (\bk)(x-x^{\prime
})}\mathcal{G}_{1}(x,x^{\prime },\bk)$ is bounded and decaying as
$\bk\rightarrow \infty $. Taking into account that the resolvent
obeys~\eqref{27} with $ U=U_{1}$ we get that the Green's function
obeys integral equations of the type~\eqref{50},
\begin{equation}
\mathcal{G}_{1}^{{}}(\bk)=\mathcal{G}_{0}^{{}}(\bk)+\mathcal{G}_{0}^{{}}(\bk)U_{1}^{{}}\mathcal{G}_{1}^{{}}(\bk
),\qquad
\mathcal{G}_{1}^{{}}(\bk)=\mathcal{G}_{0}^{{}}(\bk)+\mathcal{G}_{1}^{{}}(\bk)U_{1}^{{}}\mathcal{G}_{0}^{{}}(
\bk),  \label{171}
\end{equation}
it is analytic when $\bk_{\Re }\bk_{\Im }\neq 0$ and in this
region, in analogy with~(\ref{54}), satisfies
\begin{equation}
\frac{\partial \mathcal{G}_{1}^{{}}(\bk)}{\partial \bk_{\Im
}^{{}}}=\frac{\sgn\bk_{\Im }^{{}}}{2\pi }\,t_{1}(\bk)\varphi
(\bk)\otimes \psi (\bk).  \label{172}
\end{equation}
The discontinuity across the imaginary axis is equal to
\begin{equation}
\mathcal{G}_{1}^{{}}(+0+i\bk_{\Im
}^{{}})-\mathcal{G}_{1}^{{}}(-0+i\bk_{\Im }^{{}})=-\sgn\bk _{\Im
}^{{}}\sum_{j}t_{j}\theta (\varkappa _{j}-|\bk_{\Im
}^{{}}|)\varphi (i\varkappa _{j})\otimes \psi (i\varkappa _{j}).
\label{173}
\end{equation}
Decomposition~(\ref{164}) also gives
\begin{equation}
\mathcal{G}_{1}^{{}}(\bk)=\mathcal{G}_{1,\text{reg}}^{{}}(\bk)+\sum_{j}\gamma
_{j}(\bk)\varphi (i\varkappa _{j})\otimes \psi (i\varkappa _{j}),
\label{174}
\end{equation}
where
\begin{equation}
\gamma _{j}(\bk)=\Gamma _{j}(\ell _{\Im }^{{}}(\bk)),  \label{175}
\end{equation}
so that by~(\ref{165})
\begin{equation}
\gamma _{j}(\bk)=\frac{t_{j}\sgn\bk_{\Im }}{2\pi i}\log
\frac{\bk-i\varkappa _{j}}{\bk+i\varkappa _{j}}.  \label{176}
\end{equation}
This proves that the Green's function, in addition to the standard
discontinuity at the real axis, has also a discontinuity at the
imaginary
axis when $|\bk_{\Im }|<\max_{j}\varkappa _{j}$. Inside the quadrants $\bk%
_{\Re }\bk_{\Im }\neq 0$ the Green's function is continuous up to
the borders. Note that since
$\overrightarrow{\mathcal{L}}_{1}\varphi (i\varkappa _{j})=0$, the
regular part $\mathcal{G}_{1,\text{reg}}^{{}}(\bk)$ is still a
Green's function.

As far as the Jost solutions $\varphi $ and $\psi $ are concerned,
the integrals in definitions~(\ref{40}) are diverging, as
expected. They can be conveniently regularized for $\bk_{\Im }\neq
0$ by means of the following limiting procedure
\begin{align}
\varphi (x,\bk)& =\lim_{\varepsilon \rightarrow +0}\int dx^{\prime
}\,\bigl(
\mathcal{G}_{1}^{{}}(\bk)\overleftarrow{\mathcal{L}}_{0}^{{}}\bigr)(x,x^{\prime
})e_{{}}^{-i\ell (\bk)x^{\prime }+i\varepsilon \bk_{\Im
}^{{}}x_{2}^{\prime
}},  \label{183} \\
\psi (x,\bk)& =\lim_{\varepsilon \rightarrow +0}\int dx^{\prime
}\,e_{{}}^{i\ell (\bk)x^{\prime }-i\varepsilon \bk_{\Im
}^{{}}x_{2}^{\prime
}}\bigl(\overrightarrow{\mathcal{L}}_{0}^{{}}\mathcal{G}_{1}^{{}}(\bk)\bigr)(x^{\prime
},x).  \label{184}
\end{align}

From~(\ref{41}) we get the discontinuity of the resolvent with respect to $%
q_{2}$ when $q_{1}\neq 0$. Denoting the corresponding boundary
values of the resolvent as
\begin{equation}
\lim_{q_{2}\rightarrow \pm 0\bk_{\Im }}\left(
\widehat{M}_{1}^{{}}(q)\Bigr |_{q_{1}=\bk_{\Im }}^{{}}\right)
=\mathcal{G}_{1}^{{}}(\pm 0+i\bk_{\Im }^{{}}), \label{185}
\end{equation}
we get from~(\ref{170})
\begin{align}
\mathcal{G}_{1}^{{}}(x,x^{\prime },& \pm 0+i\bk_{\Im
})=\frac{1}{2\pi i}\int d\alpha \, \bigl[\theta
(x_{2}^{{}}-x_{2}^{\prime })-\theta (\bk_{\Im }^{{}}\alpha )
\bigr]\,t_{1}(\alpha +i\bk_{\Im })\times   \nonumber \\
& \qquad \qquad \qquad \times \varphi (x,\alpha +i\bk_{\Im })\psi
(x^{\prime
},\alpha +i\bk_{\Im })- \\
& -\sum_{j}t_{j}\theta (\varkappa _{j}-|\bk_{\Im
}^{{}}|)\,\bigl[\theta (x_{2}^{{}}-x_{2}^{\prime })-\theta (\mp
\bk_{\Im }^{{}})\bigr]\varphi (x,i\varkappa _{j})\psi (x^{\prime
},i\varkappa _{j}),  \label{186}
\end{align}
Correspondingly, decomposition~\eqref{174} gives
\begin{equation}
\mathcal{G}_{1}^{{}}(\pm 0+i\bk_{\Im
}^{{}})=\mathcal{G}_{1,\text{reg}}^{{}}(i\bk_{\Im
}^{{}})+\sum_{j}\gamma _{j}(\pm 0+i\bk_{\Im }^{{}})\,\varphi
(i\varkappa _{j})\otimes \psi (i\varkappa _{j}),  \label{187}
\end{equation}
where $\gamma _{j}(\pm 0+i\bk_{\Im }^{{}})$ is defined
by~(\ref{176}).

If we introduce the advanced/retarded Green's functions in analogy
to~(\ref {42}), by~(\ref{162}) we get the following bilinear
representation for these Green's functions in terms of the Jost
solutions on the real axis
\begin{align}
\mathcal{G}_{1,\pm }^{{}}(x,x^{\prime })=\pm \theta (\pm
(x_{2}^{{}}-x_{2}^{\prime })) \Biggl(& \frac{1}{2\pi i}\int
d\alpha \,t_{1}^{\sigma }(\alpha )\varphi ^{\sigma }(x,\alpha
)\psi ^{\sigma }(x^{\prime },\alpha )-\sum_{j}t_{j}\varphi
(x,i\varkappa _{j})\psi (x^{\prime },i\varkappa _{j})
\Biggr)  \nonumber \\
&  \label{189}
\end{align}
where $\sigma =+,-$, and, as before, we used the upper index
$\sigma $ for
the boundary values of the Jost solutions and $t_{1}(\bk)$ at the real $\bk$%
-axis. It is clear that the advanced/retarded Green's functions
are independent of the choice of $\sigma $. These Green's
functions obey the conjugation property~(\ref{57}).

The advanced/retarded solutions also must be defined by the analog
of relations~\eqref{47} of the regular case, i.e.,
\begin{align}
\varphi _{\pm }^{{}}(x,k)& =\int dx^{\prime }\,\bigl(\mathcal{G}_{1,\pm }^{{}}%
\overleftarrow{\mathcal{L}}_{0}^{{}}\bigr)(x,x^{\prime
})e_{{}}^{-i\ell
(k)x^{\prime }},  \label{190} \\
\psi _{\pm }^{{}}(x,k)& =\int dx^{\prime }\,e_{{}}^{i\ell
(k)x^{\prime }}
\bigl(\overrightarrow{\mathcal{L}}_{0}^{{}}\mathcal{G}_{1,\pm
}^{{}}\bigr)(x^{\prime },x).  \label{191}
\end{align}

\bigskip For the boundary values of the Green's function of the Jost
solutions at the real axis, from~(\ref{170}) we obtain
\begin{align}
\mathcal{G}_{1}^{\sigma }(x,x^{\prime },k)& =\frac{1}{2\pi i}\int
d\alpha \,[\theta (x_{2}-x_{2}^{\prime })-\theta (\sigma (\alpha
-k))]\,t_{1}^{\sigma }(\alpha )\varphi ^{\sigma }(x,\alpha )\psi
^{\sigma }(x^{\prime },\alpha )-
\nonumber \\
& -\sum_{j}t_{j}[\theta (x_{2}-x_{2}^{\prime })-\theta (-\sigma
k)]\,\varphi (x,i\varkappa _{j})\psi (x^{\prime },i\varkappa
_{j}),\quad k\in \R. \label{217}
\end{align}%
These functions are finite for all $k$ but discontinuous at $k=0$.
By~(\ref{189}) and~(\ref{217})
\begin{align}
\mathcal{G}_{1}^{\sigma }(k)-\mathcal{G}_{1,\pm }^{{}}& =\frac{\mp
1}{2\pi i}\int d\alpha \,\theta (\pm \sigma (\alpha
-k))t_{1}^{\sigma }(\alpha )\varphi ^{\sigma
}(\alpha )\otimes \psi ^{\sigma }(\alpha )\pm  \nonumber \\
& \pm \theta (\mp \sigma k)\sum_{j}t_{j}\varphi (i\varkappa
_{j})\otimes \psi (i\varkappa _{j}),\quad \sigma =+,-.
\label{218}
\end{align}

We observe that
\begin{equation}
\mathcal{G}_{1,\pm }^{{}}(x,x^{\prime })=\pm \theta (\pm
(x_{2}^{{}}-x_{2}^{\prime })) \Biggl(\frac{1}{2\pi i}\int d\alpha
\,\varphi _{\mp }(x,\alpha )\psi _{\pm }(x^{\prime },\alpha
)-\sum_{j}t_{j}\varphi (x,i\varkappa _{j})\psi (x^{\prime
},i\varkappa _{j})\Biggr),  \label{200:5}
\end{equation}
that proves independence of~\eqref{189} on the sign $\sigma $ in
the r.h.s. It also proves that the set of advanced/retarded
solutions is not complete.

The advanced/retarded solutions and the limiting values of the
Jost solutions at the real axis are related to each other by means
of the
following relations%
\begin{equation}
\varphi _{\pm }(k)=\int dp\,\varphi ^{\sigma
}(p)\,\overline{r_{\pm }^{-\sigma }(p,k)},\qquad t_{1}^{\sigma
}(k)\varphi ^{\sigma }(k)=\int dp\,\varphi _{\pm }(p)r_{\pm
}^{\sigma }(p,k),  \label{197}
\end{equation}
where
\begin{equation}
r_{\pm }^{\sigma }(p,k)=\delta (k-p)\left[ \theta (\pm \sigma
k)+\theta (\mp \sigma k)t_{1}(\sigma k)\right] +\theta (\mp \sigma
k)\delta (k+p)r_{1}^{\sigma }(k).  \label{198}
\end{equation}
Note that
\begin{equation}
\overline{r_{\pm }^{-\sigma }(p,k)}=\delta (k-p)\left[ \theta (\mp
\sigma k)+\theta (\pm \sigma k)t_{1}(\sigma k)\right] +\delta
(k+p)\theta (\pm \sigma k)\overline{r_{1}^{-\sigma }(k)},
\label{199}
\end{equation}
and%
\begin{equation}
\overline{\varphi _{\pm }(k)}=\psi _{\mp }(k),  \label{200:1}
\end{equation}
which allows us to write down the corresponding relations for dual
solutions.

Introducing alternative spectral data like in~\eqref{101}
\begin{equation}
f^{\sigma }={r_{\pm }^{-\sigma }}_{{}}^{\dag }r_{\pm }^{-\sigma },
\label{211}
\end{equation}
we have (cf.~\eqref{102})
\begin{equation}
\varphi ^{\sigma }T_{1}^{\sigma }=\varphi ^{-\sigma }f^{-\sigma
},\qquad T_{1}^{\sigma }\psi ^{\sigma }=f^{\sigma }\psi ^{-\sigma
},  \label{212}
\end{equation}
where $T_{1}^{\sigma }(p,k)=\delta (p)t_{1}^{\sigma }(k)$ and
$f^{\sigma }$ obeys properties (cf.~\eqref{103}):
\begin{equation}
{f_{{}}^{\sigma }}_{{}}^{\dag }=f_{{}}^{\sigma },\qquad
f_{{}}^{-\sigma }=T_{1}^{-\sigma }(f^{\sigma
}{)}_{{}}^{-1}T_{1}^{\sigma }.  \label{213}
\end{equation}


\section{Inverse scattering transform on nontrivial background:
two-dimensional perturbation of the one-dimensional potential}

\label{s4}

\subsection{Resolvent}

Now we start studying the operator $\Lo$~\eqref{NS} with potential
given in~ \eqref{potential}. In order to investigate the
properties of this operator we introduce its extension $L(q)$ like
in~\eqref{9} and the inverse to this extension, i.e., the
resolvent, by~\eqref{14}. Correspondingly, the resolvent $M(q)$
obeying~\eqref{14} will be defined through one of the integral
equations
\begin{equation}
M(q)=M_{1}^{{}}(q)+M_{1}^{{}}(q)U_{2}^{{}}M(q),\qquad
M(q)=M_{1}^{{}}(q)+M(q)U_{2}^{{}}M_{1}^{{}}(q),  \label{221}
\end{equation}%
where $U_{2}(x,x^{\prime };q)=u_{2}(x)\delta (x-x^{\prime })$. We
assume $ u_{2}(x)$ to be real, smooth, and rapidly decaying with
respect to both variables $(x_{1},x_{2})$. Moreover we assume it
to be ``small'' in the sense that solutions $M(x,x^{\prime };q)$
of the both equations in~\eqref{221} exist in $\mathcal{S}^{\prime
}(\R^{6})$. Their properties with respect to variables $q$ are
inherited from the corresponding properties of the resolvent
$M_{1}(q)$. For instance, $M(q)$ is a continuous function of $q $
when $q\neq 0$ and $q_{2}\neq 0$. The effective tool for
investifating the properties of the resolvent is given by the
Hilbert identity~\eqref{29}, which we can also write in the form
\begin{equation}
M^{\prime }(\q^{\prime })-M(\q)=M^{\prime }(\q^{\prime
})L_{1}^{\prime }(\q ^{\prime })(M_{1}^{\prime }(\q^{\prime
})-M_{1}^{{}}(\q))L_{1}^{{}}(\q)M(\q), \label{222}
\end{equation}
so that in analogy with~\eqref{32} we get for the derivatives of
the hat-kernel~\eqref{7} of the resolvent
\[
\frac{\partial \widehat{M}(q)}{\partial q_{j}^{{}}}=\widehat{M}(q)
\overleftarrow{\mathcal{L}}_{1}^{{}}\frac{\partial
\widehat{M}_{1}^{{}}(q)}{
\partial q_{j}^{{}}}\overrightarrow{\mathcal{L}}_{1}^{{}}\widehat{M}
(q),\quad j=1,2,\quad q_{2}^{{}}\neq 0.
\]
Then using~\eqref{169} we get at $q_{2}\neq 0$ equalities of the
kind~\eqref{33}, \eqref{34} for derivatives of the
$\widehat{M}(q)$:
\begin{align}
\frac{\partial \widehat{M}(q)}{\partial q_{1}^{{}}}=& \frac{i}{\pi
} \int\limits_{\bk_{\Im }=q_{1}}d\bk_{\Re }^{{}}\,\bar{\bk}\delta
(\ell _{2\Im }^{{}}(\bk)-q_{2}^{{}})\,t_{1}(\bk)\Phi (\bk)\otimes
\Psi (\bk),  \label{223}
\\
\frac{\partial \widehat{M}(q)}{\partial q_{2}^{{}}}=&
\frac{1}{2\pi i} \int\limits_{\bk_{\Im }=q_{1}}d\bk_{\Re
}^{{}}\,\delta (\ell _{2\Im }^{{}}(
\bk)-q_{2}^{{}})\,t_{1}(\bk)\Phi (\bk)\otimes \Psi (\bk),
\label{224}
\end{align}
while now the Jost solutions are defined as (cf.~(\ref{40}))
\begin{equation}
\Phi
(\bk)=\mathcal{G}(\bk)\overleftarrow{\mathcal{L}}_{1}^{{}}\varphi
(\bk),\qquad \Psi (\bk)=\psi
(\bk)\overrightarrow{\mathcal{L}}_{1}^{{}}\mathcal{G}(\bk),
\label{226}
\end{equation}
where the Green's function $\mathcal{G}(x,x^{\prime },\bk)$ of the
Jost solutions is defined as value of the resolvent by~\eqref{41}
in the same way as in the decaying case.

The resolvent is discontinuous at $q_{2}=0,$ as inherited by the
same properties of $M_{1}(q)$. Hence, we will consider separately
the cases $ q_{1}=0$ and $q_{1}\neq 0$. The boundary values of the
resolvent in the first case, $\lim_{q_{2}\rightarrow \pm
0}\widehat{M}(q)|_{q_{1}=0}$, define the advanced/retarded Green's
functions like in~\eqref{42}. In the case $ q_{1}\neq 0$ for the
boundary values of the resolvent thanks to~\eqref{41} we have in
analogy with~\eqref{185} that
\begin{equation}
\lim_{q_{2}\rightarrow \pm 0\bk_{\Im }^{{}}}\left( \widehat{M}
(q)\Bigr|_{q_{1}=\bk_{\Im }^{{}}}^{{}}\right) =\mathcal{G}(\pm
0+i\bk_{\Im }^{{}}). \label{227}
\end{equation}

\subsection{The Green's function.}

\label{s9}

By~(\ref{221}) and definition~(\ref{41}) this Green's function
satisfies integral equations
\begin{equation}
\mathcal{G}(\bk)=\mathcal{G}_{1}^{{}}(\bk)+\mathcal{G}_{1}^{{}}(\bk)U_{2}^{{}}\mathcal{G}(\bk),\qquad
\mathcal{G}(\bk
)=\mathcal{G}_{1}^{{}}(\bk)+\mathcal{G}(\bk)U_{2}^{{}}\mathcal{G}_{1}^{{}}(\bk).
\label{228}
\end{equation}
Taking into account that $\Lo=\Lo_{1}-U_{2}$, one check that
$\mathcal{G}(\bk)$
satisfies the differential equations~$\overrightarrow{\mathcal{L}}\mathcal{G}(\bk)=\mathcal{G}(%
\bk)\overleftarrow{\mathcal{L}}_{{}}^{{}}=I$ and the conjugation
property~\eqref{49}. From~\eqref{221}, \eqref{228}, it is clear
that the analyticity properties of the Green's function
$\mathcal{G}(\bk)$ are inherited from $\mathcal{G}_{1}(\bk)$. This
means that it is analytic in the region $\bk_{\Re }\bk_{\Im }\neq
0$, i.e., it obeys~\eqref{54} in this region. The Green's function
satisfies the conjugation property~\eqref{49} and possesses the
standard cut at $\bk_{\Im }=0$ and additional cut at $\bk_{\Re
}=0$ when $|\bk_{\Im }|<\max_{j}\varkappa _{j}$. Inside the
quadrants $\bk_{\Re }\bk_{\Im }\neq 0$ $\mathcal{G}(\bk)$ is
continuous up to the borders, as follows from the properties of $
\mathcal{G}_{1}(\bk)$. In particular, note that
\begin{equation}
\lim_{\sigma \bk_{\Im }\rightarrow +0}\mathcal{G}(\pm 0+i\bk_{\Im
}^{{}})=\lim_{\pm k\rightarrow +0}\mathcal{G}_{{}}^{\sigma
}(k)\equiv \mathcal{G}^{\sigma }(\pm 0),\qquad \sigma =+,-.
\label{229}
\end{equation}

As always in order to study properties of the resolvent at points
of discontinuity we use the Hilbert identity. By~\eqref{227} this
can be written as discontinuity of the Green's function across the
imaginary axis:
\begin{align}
\mathcal{G}(+0+i\bk_{\Im }^{{}})-& \mathcal{G}(-0+i\bk_{\Im
}^{{}})=-\sgn\bk_{\Im
}\sum_{j}t_{j}\theta (\varkappa _{j}-|\bk_{\Im }^{{}}|)\times   \nonumber \\
& \times \left( \mathcal{G}(\pm 0+i\bk_{\Im
}^{{}})\overleftarrow{\mathcal{L}} _{1}^{{}}\varphi (i\varkappa
_{j})\right) \otimes \left( \psi (i\varkappa
_{j})\overrightarrow{\mathcal{L}}_{1}^{{}}\mathcal{G}(\mp
0+i\bk_{\Im }^{{}}))\right) . \label{230}
\end{align}
Eq.~\eqref{230} suggests introducing the functions $\Phi
_{j}(x,\bk_{\Im })$ and $\Psi _{j}(x,\bk_{\Im })$ by means of
\begin{equation}
\Phi _{j}(\bk_{\Im }^{{}})=\mathcal{G}(+0+i\bk_{\Im
}^{{}})\overleftarrow{\mathcal{L}} _{1}^{{}}\varphi (i\varkappa
_{j}),\qquad \Psi _{j}(\bk_{\Im }^{{}})=\psi (i\varkappa
_{j})\overrightarrow{\mathcal{L}}_{1}^{{}}\mathcal{G}(+0+i\bk_{\Im
}^{{}}), \label{239}
\end{equation}
that thanks to~\eqref{154} obey
\begin{equation}
\overline{\Phi _{j}(x,\bk_{\Im })}=b_{j}\Psi _{j}(x,-\bk_{\Im
}),\qquad \overline{\Psi _{j}(x,\bk_{\Im })}=\dfrac{\Phi
_{j}(x,-\bk_{\Im })}{b_{j}}. \label{240}
\end{equation}
Now by~\eqref{230} (bottom sign) we derive for the discontinuity
the following expression
\begin{equation}
\mathcal{G}(+0+i\bk_{\Im }^{{}})-\mathcal{G}(-0+i\bk_{\Im
}^{{}})=-\sum_{l,m}\theta (\varkappa _{l}-|\bk_{\Im }^{{}}|)\Phi
_{l}(\bk_{\Im }^{{}})(A(\bk_{\Im }^{{}})^{-1})_{lm}\otimes \Psi
_{m}(\bk_{\Im }^{{}})  \label{242}
\end{equation}
where we introduced the matrix $A(\bk_{\Im }^{{}})$ with elements
\begin{equation}
A_{lm}(\bk_{\Im }^{{}})=\dfrac{\delta _{lm}}{t_{l}\sgn\bk_{\Im
}}-\theta (\min \{\varkappa _{l},\varkappa _{m}\}-|\bk_{\Im
}|)\left( \psi (i\varkappa
_{l})\overrightarrow{\mathcal{L}}_{1}^{{}}\mathcal{G}(+0+i\bk_{\Im
}^{{}}) \overleftarrow{\mathcal{L}}_{1}^{{}}\varphi (i\varkappa
_{m})\right) . \label{232}
\end{equation}
Therefore the discontinuity of the Green's function at the
imaginary axis is given not in terms of Jost solutions but in
terms of functions $\Phi _{j}(x, \bk_{\Im })$ and $\Psi
_{j}(x,\bk_{\Im })$, which are solutions of the nonstationary
Schr\"{o}dinger equation and its dual different from the Jost
ones.

One can prove that, under the assumption of unique solvability of
the integral equation for the Green's function, the matrix
$A(\bk_{\Im }^)$ is non-singular. Also, one can easily show that
\begin{equation}
A(\bk_{\Im }^{{}})^{\dag }=BA(-\bk_{\Im }^{{}})B^{-1},
\label{235}
\end{equation}%
where we introduced the diagonal matrix $B=$diag$\{b_{1},\ldots
,b_{N}\}.$ Note that $B$ is Hermitian, since the $b_{j}$'s in
(\ref{116}) are real. Preserving for the limiting values at
$\bk_{\Im }=\pm 0$ of the matrix the usual notation, $A^{\pm
}=\lim_{\bk_{\Im }^{{}}\rightarrow \pm 0}A(\bk_{\Im }^{{}})$, we
get by~\eqref{229} and~\eqref{232}
\begin{equation}
A_{lm}^{\sigma }=\dfrac{\sigma \delta _{lm}}{t_{l}}-\left( \psi
(i\varkappa
_{l})\overrightarrow{\mathcal{L}}_{1}^{{}}\mathcal{G}^{\sigma
}(+0)\overleftarrow{ \mathcal{L}}_{1}^{{}}\varphi (i\varkappa
_{m})\right) .  \label{237}
\end{equation}
These constant matrices are invertible and, due~to \eqref{235},
satisfies the conjugation property
\begin{equation}
(A_{{}}^{\sigma})^{\dag }=BA_{{}}^{-\sigma}B^{-1}, \qquad
\sigma=+,-. \label{238}
\end{equation}


\subsection{The Jost solutions.}

\label{s5}

The functions $\Phi (x,\bk)$ and $\Psi (x,\bk)$ introduced in
\eqref{226} are the Jost solutions of the nonstationary
Schr\"{o}dinger equations and its dual with
potential~\eqref{potential}, satisfying the integral equations
\begin{equation}
\Phi (\bk)=\varphi (\bk)+\mathcal{G}_{1}^{{}}(\bk)U_{2}^{{}}\Phi
(\bk),\quad \Psi (\bk)=\psi (\bk)+\Psi
(\bk)U_{2}^{{}}\mathcal{G}_{1}^{{}}(\bk). \label{261}
\end{equation}

Thanks to the properties of the Green's function
$\mathcal{G}(\bk),$ the Jost solutions are analytic functions of
$\bk\in \C$ when $\bk_{\Re }\bk _{\Im }\neq 0$ and in the generic
situation they have the standard discontinuity at the real axis,
$\bk_{\Im }=0$, and an additional discontinuity at the segment of
the imaginary axis: $\bk_{\Re }=0$, $|\bk _{\Im }|\leq
\max_{j}\varkappa _{j}$. Let us consider $|\bk_{\Im }|\neq
\varkappa _{j}$ for any $j$. The functions $\varphi (\bk)$ and
$\psi (\bk)$ are continuous at $\bk_{\Re }=0$, while the Jost
solution $\Phi (\bk)$ is not and the discontinuity is given by
\begin{equation}
\Phi (x,+0+i\bk_{\Im }^{{}})-\Phi (x,-0+i\bk_{\Im
}^{{}})=\sum_{l}\theta (\varkappa _{l}^{{}}-|\bk_{\Im }^{{}}|)\Phi
_{l}^{{}}(x,\bk_{\Im }^{{}})w_{l}^{{}}(\bk_{\Im }^{{}}),
\label{262}
\end{equation}
where we introduced functions $w_{l}(\bk_{\Im })$ as
\begin{equation}
w_{l}^{{}}(\bk_{\Im }^{{}})\equiv t_{l}^{{}}\sgn\bk_{\Im
}^{{}}\left( \psi (i\varkappa
_{l}^{{}})\overrightarrow{\mathcal{L}}_{1}^{{}}\mathcal{G}(-0+i
\bk_{\Im }^{{}})\overleftarrow{\mathcal{L}}_{1}\varphi (i\bk_{\Im
}^{{}})\right) .  \label{267}
\end{equation}
Analogously, the discontinuity of the Jost solution $\Psi (\bk)$
of the dual equation is given by
\begin{equation}
\Psi (x,+0+i\bk_{\Im }^{{}})-\Psi (x,-0+i\bk_{\Im
}^{{}})=\sum_{l}\theta (\varkappa _{l}-|\bk_{\Im }^{{}}|)\Psi
_{l}^{{}}(x,\bk_{\Im }^{{}})\overline{ w_{l}^{{}}(-\bk_{\Im
}^{{}})}.  \label{269}
\end{equation}
Note that
\begin{equation}
\overline{w_{l}^{{}}(-\bk_{\Im
}^{{}})}=\dfrac{t_{l}}{b_{l}}\sgn\bk_{\Im }^{{}}\theta (\varkappa
_{l}^{{}}-|\bk_{\Im }^{{}}|)\left( \psi (i\bk_{\Im
}^{{}})\overrightarrow{\mathcal{L}}_{1}^{{}}\mathcal{G}(-0+i\bk_{\Im
}^{{}}) \overleftarrow{\mathcal{L}}_{1}\varphi (i\varkappa
_{l}^{{}})\right) . \label{268}
\end{equation}
Thus we see by~\eqref{262} that the discontinuity of the Jost
solution across the imaginary axis is given in terms of the
functions $\Phi _{l}(x,\bk _{\Im })$, $\Psi _{l}(x,\bk_{\Im })$
introduced in~\eqref{239}. In the same way as for the Jost
solutions we prove that these functions also satisfy the
differential equations $\overrightarrow{\mathcal{L}}\Phi
_{l}^{{}}(\bk_{\Im }^{{}})=\Psi _{l}^{{}}(\bk_{\Im
}^{{}})\overleftarrow{\mathcal{L}}=0$ that follows also
from~\eqref{262}, \eqref{269}. They generalize functions $ \varphi
(i\kap_{l})$ and $\psi (i\kap_{l})$ for the case $u_{2}\neq 0$.
This results in a nontrivial dependence of $\Phi _{l}$ and $\Psi
_{l}$ on $\bk _{\Im }$. By~(\ref{230}) we need these functions on
the interval $|\bk_{\Im }|<\varkappa _{l}$ only. In what follows
we call them the auxiliary Jost solutions. As a consequence of the
properties of the total Green's function, these solutions are
discontinuous at $\bk_{\Im }=0$. Their conjugation property is
given in~\eqref{240}.

\subsubsection{Behavior of the Green's function, Jost solutions, and
spectral data at the points $\pm i\kap_j$}

\label{s16}

As it was shown in~\eqref{164} and~\eqref{174} both $M_{1}(q)$ and
$\mathcal{G}_{1}(\bk)$ have logarithmic singularities at all
points $q=(\pm \kap_{j},0) $, or correspondingly $\bk=\pm
i\kap_{j}$, $j=1,\ldots ,N$. It is clear that these singularities
affect the Green's function $\mathcal{G}(\bk)$ as well as Jost
solutions, auxiliary Jost solutions and spectral data.

Let us choose some $\kap_{j}$ and $\bk$ belongs to some
neighborhood of a point $i\kap_{j}$ or of a point $-i\kap_{j}$
that does not include other points $\pm i\kap_{l}$, $l\neq j$, or
points on the real axis. Then we introduce
\begin{equation}
\mathcal{G}_{1,j}(\bk)=\mathcal{G}_{1}(\bk)-\gamma
_{j}(\bk)\varphi (\bk)\otimes \psi (\bk),  \label{272:1}
\end{equation}
which, thanks to~\eqref{174} is finite in the vicinities of $\pm
i\kap_{j}$, while can be discontinuous there in correspondence
to~\eqref{173}. This regularization is such that$\ e^{i\ell
(\bk)(x-x^{\prime })}\mathcal{G} _{1,j}(x,x^{\prime },\bk)$ is
bounded on the $x$-plane. Now we define a new Green's function of
the nonstationary Schr\"{o}dinger equation with
potential~\eqref{potential} by means of one of the integral
equations,
\begin{align}
\mathcal{G}_{j}^{{}}(\bk)& =\mathcal{G}_{1,j}(\bk)+\mathcal{G}
_{1,j}(\bk)U_{2}\mathcal{G}_{j}(\bk),  \label{245} \\
\mathcal{G}_{j}(\bk)& =\mathcal{G}_{1,j}(\bk)+\mathcal{G}
_{j}(\bk)U_{2}\mathcal{G}_{1,j}(\bk). \label{245.1}
\end{align}
Thus $\mathcal{G}_{j}(\bk)$ is a regularization of
$\mathcal{G}(\bk)$ in the neighborhood of $\pm i\kap_{j}$ and is
such that
\begin{equation}
\mathcal{G}(\bk)=\mathcal{G}_{j}(\bk)+\gamma
_{j}(\bk)\,\widetilde{\Phi} _{j}(\bk)\otimes \Psi (\bk),
\label{246}
\end{equation}
where we use notations~\eqref{226} for the Jost solutions and
introduce in analogy
\begin{equation}
\widetilde{\Phi}
_{j}(\bk)=\mathcal{G}_{j}(\bk)\overleftarrow{\mathcal{L}}
_{1}\varphi (\bk),\qquad \widetilde{\Psi} _{j}(\bk)=\psi (\bk)
\overrightarrow{\mathcal{L}}_{1}\mathcal{G}_{j}(\bk). \label{247}
\end{equation}
The functions $\tilde{\Phi} _{j}(\bk)$ and $\tilde{\Psi}
_{j}(\bk)$ are also solutions of the nonstationary Schr\"{o}dinger
equation and its dual with potential~\eqref{potential}. These
functions are bounded in vicinities of the points $\pm i\kap_{j}$,
though their limits at these points can be dependent on the sign
of $\bk_{\Re }$. The same properties hold for function
\begin{equation}
g_{j}(\bk)=\left( \psi
(\bk)\overrightarrow{\mathcal{L}}_{1}^{{}}\mathcal{G}
_{j}(\bk)\overleftarrow{\mathcal{L}}_{1}\varphi (\bk)\right).
\label{247.2}
\end{equation}
Thus we see that behavior of the Green's function and of the Jost
solutions in vicinities of points $\pm i\kap_{j}$ is determined by
behavior of the function $g_{j}(\bk)$ at these points. If
\begin{equation}
g_{j}(\bk)=o(1),\qquad \bk\sim i\kap_{j},  \label{255:1}
\end{equation}
the Green's function has logarithmic singularities at points $\pm
i \kap_{j}$ and the Jost solutions are bounded at these points and
their limits depend on the sign of $\bk_{\Re }.$ On the other
hand, if
\begin{equation}
g_{j}(\bk)=O(1),\qquad \bk\sim i\kap_{j},  \label{255}
\end{equation}
and the limit (also depending on the way of the limiting
procedure) is different from zero, then
\begin{equation}
\mathcal{G}(\bk)=\mathcal{G}_{j}(\bk) -\frac{1}{g_{j}(\bk)}
\tilde{\Phi}_{j}(\bk)\otimes \tilde{\Psi}_{j}(\bk) ,\qquad \bk\sim
\pm i\kap_{j}, \label{257.1}
\end{equation}
and
\begin{equation}
\Phi (\bk)=o(1),\qquad \Psi (\bk)=o(1),\qquad \bk\sim \pm
i\varkappa _{j}, \label{275}
\end{equation}
and in the r.h.s.'s we have terms of order $1/\log $.

In what follows we assume that condition~\eqref{255} is fulfilled
for all $ j=1,\ldots ,N.$ In particular, under such an assumption,
one can show that
\begin{equation}
\Phi _{j}(\pm \varkappa _{j})=\Psi _{j}(\pm \varkappa
_{j})=0,\qquad j=1,\ldots ,N  \label{279}
\end{equation}
while the values $\Phi _{m}(\pm \varkappa _{j})$ and $\Psi
_{m}(\pm \varkappa _{j})$ of the auxiliary solutions for $m$ such
that $\kap_{m}>\kap _{l}$ are finite and different from zero, and
\begin{equation}
\psi (i\kap_{j})\overrightarrow{\mathcal{L}}_{1}\Phi _{m}(\pm
\varkappa _{j})=0,\qquad \Psi _{m}(\pm \varkappa
_{j})\overleftarrow{\mathcal{L}} _{1}\varphi (i\kap_{j})=0.
\label{279:1}
\end{equation}
Taking into account~\eqref{247} and~\eqref{279:1} one gets
\begin{equation}
A(\pm \kap_{j})_{jm}=A(\pm \kap_{j})_{mj}=\dfrac{\pm \delta
_{jm}}{t_{j}}, \label{253:1}
\end{equation}
that coincide with values of $A_{jm}(\bk_{\Im })$ and
$A_{jm}(\bk_{\Im })$ when $\bk_{\Im }>\min \{\varkappa
_{j},\varkappa _{m}\}$. In the same way we derive
\begin{equation}
A(\pm \kap_{j})_{jm}^{-1}=A(\pm \kap_{j})_{mj}^{-1}=\pm \delta
_{jm}t_{j}. \label{254}
\end{equation}
All other matrix elements of matrices $A(\bk_{\Im })$ and
$A^{-1}(\bk_{\Im }) $ have finite limits at $\bk_{\Im }=\pm
\kap_{j}$.

Finally, one can prove that under condition~\eqref{255}
\begin{equation}
w_{l}^{{}}(\pm \kap_{j})_{jm}=0\quad \text{for all $j$ and
$l$:}\,\kap _{j}\leq \kap_{l}.  \label{284}
\end{equation}

The properties of the Jost solutions and spectral data derived so
far enable us to reconstruct the auxiliary Jost solutions in terms
of the boundary values of the Jost ones. Indeed, the knowledge of
derivatives of Green's function and auxiliary Jost solutions with
respect to $\bk_{\Im }^{{}}$ allows to obtain
\begin{equation}
\mathcal{G}(\pm 0+i\bk_{\Im }^{{}})=\mathcal{G}_{{}}^{\sigma }(\pm
0)+\frac{ \sigma }{2\pi }\int\limits_{0}^{\bk_{\Im }}d\alpha
\,t_{1}(i\alpha )\Phi (\pm 0+i\alpha )\otimes \Psi (\pm 0+i\alpha
),\quad \sigma =\sgn\bk_{\Im }^{{}},  \label{244}
\end{equation}%
\begin{equation}
\theta (\varkappa _{j}-|\mathbf{k}_{\Im }|)\Phi
_{j}(\mathbf{k}_{\Im })= \frac{\sgn\mathbf{k}_{\Im }}{2\pi
}\int_{\varkappa _{j}\sgn\mathbf{k}_{\Im }}^{\mathbf{k}_{\Im
}}d\alpha \ t_{1}(i\alpha )\Phi (+0+i\alpha
)\sum_{l}b_{l}\overline{w_{l}(-\alpha )}A_{lj}(\alpha ).
\label{280}
\end{equation}
Also, from the expression for the derivative of
$A_{mj}(\mathbf{k}_{\Im })$ one gets
\begin{align}
\left( A^{-1}(\mathbf{k}_{\Im })\right) _{jm}=&
t_{m}\sgn\mathbf{k}_{\Im }\delta _{jm}+\theta (\min \{\varkappa
_{m},\varkappa _{j}\}-|\mathbf{k}
_{\Im }|)\frac{\sgn\mathbf{k}_{\Im }}{2\pi }b_{m}\times  \nonumber \\
& \times \int\limits_{\min \{\varkappa _{m},\varkappa
_{j}\}\sgn\mathbf{k} _{\Im }}^{\mathbf{k}_{\Im }}\!\!\!d\alpha
\,t_{1}(i\alpha )w_{j}(\alpha ) \overline{w_{m}(-\alpha )}.
\label{282}
\end{align}
It is straightforward to verify that the properties of the Jost
solutions and spectral data described above are compatible with
representations~\eqref{280} and~\eqref{282}.

Finally, one can derive bilinear representation of the resolvent
in terms of the Jost solutions
\begin{align}
& \widehat{M}(x,x^{\prime };q)=\frac{1}{2\pi
i}\int\limits_{\bk_{\Im }=q_{1}}d\bk_{\Re }\,\bigl[\theta
(x_{2}^{{}}-x_{2}^{\prime })-\theta (2\bk _{\Re }^{{}}\bk_{\Im
}^{{}}-q_{2}^{{}})\bigr]\,t_{1}(\bk)\Phi (x,\bk)\Psi
(x^{\prime },\bk)-  \nonumber \\
& -\sgn q_{1}\,\bigl[\theta (x_{2}^{{}}-x_{2}^{\prime })-\theta
(-q_{2}^{{}}) \bigr]\sum_{l,m}\theta (\min \{\varkappa
_{l},\varkappa _{m}\}-|q_{1}|)\bigl( A(q_{1})^{-1}\bigr)_{lm}\Phi
_{l}(x,q_{1})\Psi _{m}(x^{\prime },q_{1}). \label{285}
\end{align}
Note that the corresponding kernel $M(x,x^{\prime };q)$ defined
by~\eqref{7} belongs to $\Ss^{\prime }$. Continuity of the
resolvent at $q_{1}=0$ in the case where $q_{2}\neq 0$ implies
that
\begin{align}
& \dfrac{1}{2\pi i}\int dk\ \,t_{1}^{+}(k)\Phi _{{}}^{+}(k)\otimes
\Psi _{{}}^{+}(k)-\sum_{lm}(A^{+})_{lm}^{-1}\Phi _{l}^{+}\otimes
\Psi _{m}^{+}=
\nonumber \\
=& \dfrac{1}{2\pi i}\int dk\,\ t_{1}^{-}(k)\Phi
_{{}}^{-}(k)\otimes \Psi
_{{}}^{-}(k)+\sum_{lm}(A^{-})_{lm}^{-1}\Phi _{l}^{-}\otimes \Psi
_{m}^{-} \label{286}
\end{align}
where we introduced
\begin{equation}
\Phi _{l}^{\pm }(x)=\lim_{\pm \bk_{\Im }^{{}}\rightarrow +0}\Phi
_{l}^{{}}(x, \bk_{\Im }^{{}}),\qquad \Psi _{l}^{\pm }(x)=\lim_{\pm
\bk_{\Im }^{{}}\rightarrow +0}\Psi _{l}^{{}}(x,\bk_{\Im }^{{}}).
\label{287}
\end{equation}
Thanks to~(\ref{67}) and~(\ref{240}) these limiting values obey
the following conjugation properties:
\begin{equation}
\overline{\Phi _{{}}^{\pm }(x,k)}=\Psi _{{}}^{\mp }(x,k),\qquad
\overline{ \Phi _{l}^{\pm }(x)}=b_{l}\Psi _{l}^{\mp }(x),\quad
k\in \R,\quad l=1,\ldots ,N.  \label{288}
\end{equation}

The bilinear representation~\eqref{285} for the resolvent leads,
recalling (\ref{37}), to the following bilinear representation for
the Green's function of the Jost solutions:
\begin{align}
& \mathcal{G}(x,x^{\prime },\bk)=\frac{1}{2\pi i}\int dk^{\prime
}\,\bigl[ \theta (x_{2}^{{}}-x_{2}^{\prime })-\theta (\bk_{\Im
}^{{}}k^{\prime })\bigr] \,t_{1}^{{}}(k^{\prime }+\bk)\Phi
(x,k^{\prime }+\bk)\Psi (x^{\prime
},k^{\prime }+\bk)-  \nonumber \\
& \quad -\sgn\bk_{\Im }\bigl[\theta (x_{2}^{{}}-x_{2}^{\prime
})-\theta (-\bk_{\Re }^{{}}\bk_{\Im }^{{}})\bigr] \sum_{l,m}\theta
(\min \{\varkappa _{l},\varkappa _{m}\}-|\bk_{\Im }^{{}}|)
\bigl(A(\bk_{\Im })^{-1}\bigr)_{lm}\times  \nonumber \\
& \quad \times \Phi _{l}(x,\bk_{\Im }^{{}})\Psi _{m}(x^{\prime
},\bk_{\Im }^{{}}),  \label{289}
\end{align}
that generalizes~\eqref{170}. Below we use the bilinear
representation to obtain relations between the Jost and
advanced/retarded solutions.

\subsection{Discontinuity of the resolvent at $q=0$}

First, we introduce advanced/retarded Green's functions as
specific limits of the resolvent in analogy to~\eqref{42}. It is
straightforward to prove that they satisfy the differential
equations equations~$\overrightarrow{ \mathcal{L}}\mathcal{G}_{\pm
}=\mathcal{G}_{\pm }\overleftarrow{\mathcal{L}} =I$ and~obey
\eqref{57} and integral equations~\eqref{58}. In order to find
difference among these Green's function we use the Hilbert
identity, (cf.~(\ref{47})).

The bilinear representation~\eqref{285} for the resolvent, thanks
to~(\ref{42}), gives a representation for the advanced/retarded
Green's functions in terms of the Jost solutions on the real axis:
\begin{align}
\mathcal{G}_{\pm }^{{}}(x,x^{\prime })=\pm \theta (\pm
(x_{2}^{{}}-x_{2}^{\prime }))\Biggl(& \frac{1}{2\pi i}\int
dk\,t_{1}^{\sigma }(k)\Phi _{{}}^{\sigma }(x,k)\Psi _{{}}^{\sigma
}(x^{\prime },k)-  \nonumber
\\
& -\sigma \sum_{l,m}(A_{{}}^{\sigma })_{lm}^{-1}\Phi _{l}^{\sigma
}(x)\Psi _{m}^{\sigma }(x^{\prime })\Biggr),  \label{294}
\end{align}%
where the standard and auxiliary advanced/retarded solutions are
defined as
\begin{align}
& \Phi _{\pm }^{{}}(k)=\mathcal{G}_{\pm
}^{{}}\overleftarrow{\mathcal{L}} _{1}^{{}}\varphi _{\pm
}^{{}}(k), & & \Phi _{\pm ,j}^{{}}=\mathcal{G}_{\pm
}^{{}}\overleftarrow{\mathcal{L}}_{1}^{{}}\varphi (i\kap_{j}),
\label{291}
\\
& \Psi _{\pm }^{{}}(k)=\psi _{\pm
}^{{}}(k)\overrightarrow{\mathcal{L}} _{1}^{{}}\mathcal{G}_{\pm
}^{{}}, & & \Psi _{\pm ,j}^{{}}=\psi (i\kap_{j})
\overrightarrow{\mathcal{L}}_{1}^{{}}\mathcal{G}_{\pm }^{{}}.
\label{292}
\end{align}
These functions are solutions of the nonstationary Schr\"{o}dinger
equation with potential $u$, obey conjugation
properties~\eqref{77} and also are such that
\begin{equation}
\overline{\Phi _{\pm ,j}}=b_{j}\Psi _{\mp ,j}.  \label{293}
\end{equation}
We mention that the l.h.s. of (\ref{294})\ is independent of the
sign $ \sigma =+,-$ due to condition~\eqref{286}.

\subsubsection{Relations among the Green's functions.}

\label{s19}

Another limiting procedure at point $q=0$ for the resolvent $M(q)$
is given by the limiting values of the Green's function
$\mathcal{G}(\bk)$ at the real axis. For these boundary values we
use notation~$\mathcal{G} _{{}}^{\sigma }(k)$. The difference
between these limiting values and the advanced/retarded Green's
functions, like in the case of the decaying potential, can be
presented in two forms. The first one follows from the limits
$\bk\rightarrow k\pm i0$ in~\eqref{289} and~\eqref{294}:
\begin{align}
\mathcal{G}_{{}}^{\sigma }(k)-\mathcal{G}_{\pm }^{{}}=& \frac{\mp
1}{2\pi i} \int dk^{\prime }\,\theta (\pm \sigma (k^{\prime
}-k))\,t_{1}^{\sigma }(k^{\prime })\Phi _{{}}^{\sigma }(k^{\prime
})\otimes \Psi _{{}}^{\sigma
}(k^{\prime })\pm  \nonumber \\
& \pm \sigma \theta (\mp \sigma k)\sum_{l,m}(A_{{}}^{\sigma
})_{lm}^{-1}\Phi _{l}^{\sigma }\otimes \Psi _{m}^{\sigma }.
\label{295}
\end{align}
A second set of relations, obtained from the Hilbert identity, is
given by
\begin{align}
\mathcal{G}_{{}}^{\sigma }(k)-\mathcal{G}_{\pm }^{{}}& =\frac{\mp
1}{2\pi i} \int d\alpha \,\theta (\pm \sigma (\alpha -k))\int
dp\,\Phi _{\pm }^{{}}(p)r_{\pm }^{\sigma }(p,\alpha )\otimes \psi
^{\sigma }(\alpha )
\overrightarrow{\mathcal{L}}_{1}^{{}}\mathcal{G}_{{}}^{\sigma
}(k)\pm
\nonumber \\
& \pm \theta (\mp \sigma k)\sum_{j}t_{j}\Phi _{\pm ,j}\otimes \psi
(i\varkappa _{j})\overrightarrow{\mathcal{L}}_{1}^{{}}\mathcal{G}
_{{}}^{\sigma }(k),  \label{296}
\end{align}
and the analogous equality that can be derived from this last one
by conjugation.

\subsection{Spectral data.}

\label{s7}

From~\eqref{295} and~\eqref{296} we obtain the following relations
between the Jost and advanced/retarded solutions
\begin{align}
& t_{1}^{\sigma }(k)\Phi _{{}}^{\sigma }(k)=\int dp\,\Phi _{\pm
}^{{}}(p)\Ro _{\pm }^{\sigma }(p,k)+\sum_{j}\Phi _{\pm
,j}^{{}}\Ro_{\pm ,j}^{\sigma
}(k),\quad \sigma =+,-,  \label{298} \\
& \Phi _{m}^{\sigma }=\int dp\,\Phi _{\pm
}^{{}}(p)\widehat{\Ro}_{\pm ,m}^{\sigma }(p)+\theta (\pm \sigma
)\Phi _{\pm ,m}^{{}}\mp \theta (\mp \sigma )\sum_{j}t_{j}\Phi
_{\pm ,j}^{{}}A_{jm}^{\sigma },  \label{299}
\end{align}
where we used~\eqref{232} in the last line and where we introduced
the spectral data
\begin{equation}
\Ro_{\pm }^{\sigma }(p,k)=\left(r_{\pm }^{\sigma }R_{\pm }^{\sigma
}\right)(p,k)\equiv \int d\alpha \ r_{\pm }^{\sigma
}(p,\alpha)R_{\pm }^{\sigma }(\alpha,k), \label{300:1}
\end{equation}
where $R_{\pm }^{\sigma }$ is a triangular operator
\begin{equation}
R_{\pm }^{\sigma }(p,k)=\delta (p-k)\mp \theta (\pm \sigma
(p-k))R_{{}}^{\sigma }(p,k),  \label{300}
\end{equation}
with
\begin{equation}
R^{\sigma }(p,k)=
= t_{1}^{\sigma }(k)\frac{\psi _{{}}^{\sigma }(p)
\overrightarrow{\mathcal{L}}_{1}^{{}}\Phi _{{}}^{\sigma }(k)}{2\pi
i}\equiv t_{1}^{\sigma }(k)\frac{\psi
_{{}}^{\sigma}(p)U_{2}^{{}}\Phi _{{}}^{\sigma }(k)}{2\pi i}
\label{301}
\end{equation}
and
\begin{align}
& \Ro_{\pm ,j}^{\sigma }(k)=\pm \theta (\mp \sigma
k)\Ro_{j}^{\sigma }(k),
\label{302} \\
& \Ro_{j}^{\sigma }(k)=t_{j}^{{}}t_{1}^{\sigma }(k)\bigl(\psi
(i\varkappa
_{j})\overrightarrow{\mathcal{L}}_{1}^{{}}\mathcal{G}_{{}}^{\sigma
}(k) \overleftarrow{\mathcal{L}}_{1}^{{}}\varphi _{{}}^{\sigma
}(k)\bigr) \label{303:1}
\end{align}
and
\begin{equation}
\widehat{\Ro}_{\pm ,m}^{\sigma }(p)=\left(r_{\pm }^{\sigma
}\widehat{R}_{\pm ,m}^{\sigma }\right)(p)\equiv \int d\alpha \,
r_{\pm }^{\sigma }(p,\alpha)\widehat{R}_{\pm ,m}^{\sigma
}(\alpha), \label{304:1}
\end{equation}
with
\begin{align}
& \widehat{R}_{\pm ,m}^{\sigma }(k)=\mp \theta (\pm \sigma
k)\widehat{R}
_{m}^{\sigma }(k),  \label{304} \\
& \widehat{R}_{m}^{\sigma }(k)=
\frac{\psi
_{{}}^{\sigma}(k)\overrightarrow{\mathcal{L}}_{1}^{{}}\Phi
_{m}^{\sigma }}{2\pi i}\equiv \frac{\psi _{{}}^{\sigma
}(k)U_{2}^{{}}\Phi _{m}^{\sigma }}{2\pi i}. \label{305:1}
\end{align}
These relations combine~\eqref{88} and~\eqref{197}.
Eqs.~\eqref{298}, \eqref{299} give the boundary values of the Jost
solutions in terms of the advanced/retarded ones. Inverse
relations are given by
\begin{align}
\Phi _{\pm }^{{}}(k)=& \int dk^{\prime }\Phi _{{}}^{\sigma
}(k^{\prime }) \overline{\Ro_{\pm }^{-\sigma }(k,k^{\prime
})}-2\pi i\sigma \sum_{l,m}\Phi _{l}^{\sigma }(A_{{}}^{\sigma
})_{lm}^{-1}\frac{1}{b_{m}}\overline{\widehat{
\Ro}_{\pm ,m}^{-\sigma }(k)},  \label{311} \\
\Phi _{\pm ,j}^{{}}=& \frac{-b_{j}}{2\pi it_{j}}\int dk^{\prime
}\Phi _{{}}^{\sigma }(k^{\prime })\overline{\Ro_{\pm ,j}^{-\sigma
}(k^{\prime })}+\theta (\pm \sigma )\Phi _{j}^{\sigma }\mp
\frac{\theta (\mp \sigma )}{t_{j}}\sum_{l}\Phi _{l}^{\sigma
}(A_{{}}^{\sigma })_{lj}^{-1}. \label{312}
\end{align}

If we introduce alternative spectral data in analogy
with~\eqref{101}, \eqref{102} and~\eqref{211}, \eqref{212}, we can
write
\begin{align}
& t_{1}^{\sigma }(k)\Phi _{{}}^{\sigma }(k)=\int dp\,\Phi
_{{}}^{-\sigma }(p) \Fo_{{}}^{-\sigma }(p,k)+2\pi i\sigma
\sum_{l,m}\Phi _{l}^{-\sigma
}(A_{{}}^{-\sigma })_{lm}^{-1}\frac{1}{b_{m}}\Fo_{m}^{-\sigma }(k),\label{404} \\
& \Phi _{j}^{\sigma }=\int dp\,\Phi _{{}}^{-\sigma
}(p)\widehat{\Fo}_{j}^{-\sigma } (p)+2\pi i\sigma \sum_{l,m}\Phi
_{l}^{-\sigma }(A_{{}}^{-\sigma
})_{lm}^{-1}\frac{1}{b_{m}}\Fo_{mj}^{-\sigma }, \label{405}
\end{align}
where
\begin{align}
& \Fo_{{}}^{-\sigma }(k,k^{\prime })=\int dk^{\prime \prime }\,
\overline{\Ro_{\pm }^{\sigma }(k^{\prime \prime },k)}\Ro_{\pm
}^{\sigma }(k^{\prime \prime },k^{\prime
})-\sum_{j}\frac{b_{j}}{2\pi it_{j}}\overline{\Ro_{\pm
,j}^{\sigma }(k)}\Ro_{\pm ,j}^{\sigma }(k^{\prime }) \label{406} \\
& \Fo_{l}^{-\sigma }(k^{\prime })=-\frac{\theta (\pm \sigma
)b_{l}}{2\pi it_{l}}\Ro_{\pm ,l}^{\sigma }(k)\pm \frac{\theta (\mp
\sigma )}{2\pi i}
\sum_{m=1}^{N}b_{l}A_{lm}^{-\sigma }\Ro_{\pm ,m}^{\sigma }(k)+  \notag \\
& \quad \quad +\int dk^{\prime \prime
}\,\overline{\widehat{\Ro}_{\pm ,l}^{\sigma }(k^{\prime \prime
})}\Ro_{\pm }^{\sigma }(k^{\prime \prime
},k^{\prime }) \\
& \widehat{\Fo}_{j}^{-\sigma }(k)=-\dfrac{\theta (\pm \sigma
)b_{j}}{2\pi it_{j}}\overline{\Ro_{\pm ,j}^{\sigma }(k)}\pm
\sum_{l}\frac{b_{l}\theta (\mp \sigma )}{2\pi i}A_{lj}^{\sigma
}\overline{\Ro_{\pm ,l}^{\sigma }(k)}+
\notag \\
& \quad \quad +\int dk^{\prime }\overline{\Ro_{\pm }^{\sigma
}(k^{\prime },k)
}\,\widehat{\Ro}_{\pm ,j}^{\sigma }(k^{\prime }) \\
& \Fo_{lj}^{-\sigma }=-\dfrac{\theta (\pm \sigma )b_{l}}{2\pi
it_{l}}\delta _{lj}+\frac{\theta (\mp \sigma )}{2\pi
i}\sum_{m=0}^{N}b_{l}A_{lm}^{-\sigma }t_{m}A_{{}}^{\sigma
}{}_{mj}^{{}}+\int dk^{\prime }\, \overline{\widehat{\Ro}_{\pm
,l}^{\sigma }(k^{\prime })}\widehat{\Ro}_{\pm ,j}^{\sigma
}(k^{\prime }). \label{409}
\end{align}
Characterization equations for spectral data will be given in a
forthcoming publication.

\section{Conclusion}

\label{s8}

Existence of the bilinear representation~\eqref{81} is one of the
main advantages of the resolvent approach. This relation gives the
extended resolvent in terms of the Jost solutions. These solutions
themselves or any other relevant solutions of the linear problem
under consideration (in our case~\eqref{NS}) and its dual are
given as specific reductions of the corresponding Green's
functions. At the same time, the Green's functions, in their turn,
are values of the resolvent in some specific points. This property
supplies us with the simple and regular way of deriving relations
between different kinds of solutions of the linear problem and
construction of the scattering data.

\appendix

\end{document}